\documentclass[prd, aps, twocolumn, superscriptaddress, floatfix, nofootinbib, amsmath, amssymb, preprintnumbers]{revtex4-1}

\usepackage{graphicx}
\usepackage{dcolumn}
\usepackage{verbatim}
\usepackage{bm}
\usepackage{breakurl}
\usepackage{todonotes}
\usepackage{lmodern}

\usepackage{tikz}
\usepackage{pgfplots}
 \usepackage{amsmath,mathrsfs,graphicx,placeins,float}
 \usepackage{booktabs}
 \usepackage{multirow}
 \usepackage{pstricks}
\usepackage{subcaption}
\usepackage{graphicx} 
\usepackage{comment}
\usepackage{amsmath,amsfonts,amssymb,mathrsfs,graphicx,epstopdf,placeins,float,comment} \usepackage[T1]{fontenc} \usepackage{xcolor} \usepackage{cancel} 
\pgfplotsset{compat=newest,every axis plot/.append style={line width=1pt}}
\usepackage[colorlinks,linktocpage,linkcolor=cyan,citecolor=cyan]{hyperref}
\usepackage{tabularx}

\definecolor{lightgray}{gray}{0.9}
\definecolor{Amber}{rgb}{1.0, 0.75, 0.0}
\definecolor{blizzardblue}{rgb}{0.67, 0.9, 0.93}

\begin{document}

\title{Bipartite entanglement of the primordial Majorana during inflation}

\author{Ai-chen Li}
\email{alexkenlee@163.com}
\affiliation{Department of Physics, Institute of Fundamental Physics and Quantum Technology, Ningbo University, 818 Fenghua Road, Ningbo, 315211, Zhejiang, China}
\affiliation{School of Physics and Optoelectronic Engineering
Beijing University of Technology}
\affiliation{Departament de F\'{i}sica Qu\`{a}ntica i Astrof\'{i}sica, Institut de Ci\`{e}ncies del Cosmos (ICCUB), Universitat de Barcelona, Mart\'{i} i Franqu\`{e}s 1, E-08028 Barcelona, Spain}
\author{Han-Qing Shi}
\email{hqshi@bjut.edu.cn}
\affiliation{School of Physics and Optoelectronic Engineering
Beijing University of Technology}
\author{Keyun Wu}
\email{keyunwu@fqa.ub.edu}
\affiliation{Departament de F\'{i}sica Qu\`{a}ntica i Astrof\'{i}sica, Institut de Ci\`{e}ncies del Cosmos (ICCUB), Universitat de Barcelona, Mart\'{i} i Franqu\`{e}s 1, E-08028 Barcelona, Spain}

\begin{abstract}
We use a primordial Majorana field as a fermionic probe of quantum correlations during inflation. Working in a torsion-free FLRW spacetime, we derive the two-component Majorana mode equations in an axion-inflation background and construct the corresponding quadratic Hamiltonian in the paired momentum basis. Hamiltonian diagonalization and the fermionic squeezing formalism are shown to give the same Bogoliubov transformation, providing a direct map from the Majorana mode functions to the instantaneous occupation number and to the two-mode state of each $(\boldsymbol{k},-\boldsymbol{k})$ pair. Because Fermi statistics restricts each helicity sector to the vacuum and one-pair states, the resulting Hilbert space is finite and the bipartite quantum-information measures can be evaluated explicitly. We compute the von Neumann entropy of the reduced mode and the logarithmic negativity of the Majorana pair. Both diagnostics indicate that sufficiently light Majorana modes can retain enhanced super-horizon bipartite quantumness, with the logarithmic negativity making the residual inseparability especially explicit. Our result does not by itself constitute an observational Bell test or a complete decoherence analysis; rather, it identifies a Pauli-bounded matter sector in which horizon exit alone is not sufficient to erase the quantum signature encoded in the two-mode state, thereby motivating an open-system study of how reheating and inflaton-induced interactions classicalize primordial fermionic probes.
\end{abstract}

\maketitle

\tableofcontents

\section{Introduction}

Inflation provides a natural arena in which microscopic quantum fluctuations are stretched to cosmological scales, thereby seeding the primordial correlations observed at late times. This idea was shaped by a series of seminal realizations of accelerated expansion, including $R^2$ inflation, the original false-vacuum scenario, new inflation, and chaotic inflation \cite{Starobinsky:1980te,Guth:1980zm,Linde:1981mu,Albrecht:1982wi,Linde:1983gd}, while the quantum origin of scalar perturbations was already emphasized in early analyses of inflationary fluctuations \cite{Mukhanov:1981xt}. In parallel, the tensor sector supplied an equally important line of thought: primordial gravitons arise from the amplification of vacuum gravitational-wave fluctuations in an expanding universe \cite{Grishchuk:1974ny,Ford:1977dj,Starobinsky:1979ty,Rubakov:1982df}. These developments make inflation a powerful bridge between quantum field theory and cosmological observation, but they also leave a sharp conceptual question: in what precise sense do quantum states on super-Hubble scales become classical, and which genuinely quantum correlations survive the squeezing, coarse graining, and possible decoherence of the inflationary state? Early squeezed-state and semiclassical analyses showed that both scalar inflaton perturbations and tensor graviton modes naturally evolve into highly squeezed two-mode states with large occupation numbers and approximately classical stochastic phases \cite{Grishchuk:1990bj,Albrecht:1992kf,Polarski:1995jg,Kiefer:1998qe}. However, these arguments largely amount to classical behavior inferred from kinematics or reduced descriptions: without specifying a physical environment, a pointer basis, and an operational criterion for coherence loss, they do not by themselves provide a complete dynamical account of decoherence. Subsequent work sharpened this limitation as a version of the cosmological measurement problem \cite{Martin:2012pea}. A complementary operational perspective asks whether primordial correlations could, at least in principle, be certified by Bell-type or other non-classicality witnesses. The explicit construction of cosmological Bell inequalities makes this question concrete: it shows how inflationary correlations may be phrased in terms of quantum witnesses, while also highlighting that the existence of entanglement, its observational accessibility, and its survival under decoherence are distinct issues \cite{Maldacena:2015bha}. The decoherence program then moved toward more explicit interaction-based mechanisms, including environmental decoherence of primordial fluctuations, gravitational nonlinearities, nonlinear scalar-tensor couplings, non-Gaussian signatures of decoherence, and minimal decoherence from unavoidable inflationary interactions \cite{Burgess:2006jn,Nelson:2016kjm,Nelson:2017pmc,Martin:2018lin,Ye:2018kty,Burgess:2022nwu,Sou:2022nsd}. At the same time, critical analyses have stressed that intrinsic decoherence is not automatic \cite{Hsiang:2021kgh}, while quantum-information diagnostics have been used to track entanglement directly in inflationary settings \cite{Brahma:2020zpk,Brahma:2024ycc}. Taken together, these works show that scalar inflaton modes and tensor gravitons remain the standard testing ground for inflationary classicalization, but they also make clear that the quantum versus classical nature of super-horizon perturbations is still an open problem rather than a settled conclusion. This motivates asking whether complementary probes can display residual quantum correlations on super-Hubble scales in a cleaner and more diagnostic way.

Beyond the inflaton and graviton, inflation can also excite spectator scalars, gauge fields, and fermionic fields. Such additional sectors are not merely model-building decorations: they can carry spin, chirality, mass thresholds near the Hubble scale, and distinct statistics, and therefore probe aspects of the quantum state that are hidden in purely bosonic scalar or tensor perturbations. In this work we revisit the quantum-to-classical question using a fermionic probe rather than the inflaton perturbation itself. Fermion-type fields are intrinsically quantum objects in a particularly sharp sense: their excitations obey anti-commutation relations, their occupation numbers are Pauli-limited, and their states cannot be approximated by arbitrarily large classical field amplitudes in the same way as bosonic squeezed states. This makes them especially persuasive probes of genuine quantumness, compared with the inflaton sector where the meaning of residual quantum coherence after horizon exit remains conceptually debated. A Majorana field is useful for this purpose \cite{Majorana:1937vz}. For each momentum pair $(\mathbf{k},-\mathbf{k})$ and helicity sector, the Grassmann nature of the field and the Pauli principle restrict the relevant state to a finite two-mode Hilbert space, so that particle production, squeezing, and entanglement are directly tied to a bounded occupation number. This simplicity is not only physical but also computational: the fermionic squeezing operator truncates, the two-mode state has an elementary $\alpha_k |0_{\mathbf{k}},0_{-\mathbf{k}}\rangle+\beta_k|1_{\mathbf{k}},1_{-\mathbf{k}}\rangle$ structure, and quantum-information measures that are difficult to evaluate for bosonic squeezed states can often be computed explicitly. Fermions produced during axion inflation \cite{Adshead:2015kza,Adshead:2015jza,Adshead:2018oaa} therefore provide a clean setting in which to ask whether super-horizon modes are merely effectively classical, or whether they retain measurable bipartite quantumness. Motivated by recent quantum-information studies of cosmological perturbations during inflation, including entanglement entropy, purity, quantum discord, decoherence bounds, and logarithmic negativity \cite{Lim:2014uea,Martin:2015qta,Brahma:2020zpk,Martin:2021znx,Brahma:2024ycc,Liu:2026mzz,Wang:2025hlz,Sun:2019obt,Chen:2016xqz,Chen:2017cgw}, we quantify this question for primordial Majorana modes by computing the von Neumann entropy and logarithmic negativity of the corresponding two-mode state, using two-component spinor conventions \cite{Dreiner:2008tw}.

Fermionic quantum-information measures have already led to useful lessons in de Sitter spacetime. The Dirac-field vacuum in de Sitter space exhibits nontrivial entanglement entropy, with a dependence on mass, momentum, and the choice of field-theoretic bipartition \cite{Kanno:2016qcc}. Subsequent studies of fermionic entanglement measures in the cosmological de Sitter background showed that entanglement entropy, logarithmic negativity, and related correlation diagnostics can respond sensitively to fermion mass and spacetime expansion \cite{Bhattacharya:2020bal}. Fermionic Bell-type violations in de Sitter space, especially in the presence of background electromagnetic fields, further indicate that spinor probes can retain non-classical correlations in a way that is not simply captured by scalar squeezed-state intuition \cite{Ali:2021jch}. These results suggest that fermions are not only technically tractable probes, but also conceptually sharp diagnostics of primordial quantumness. We therefore ask whether analogous fermionic quantum-information signals persist in a more realistic inflationary production mechanism, and in particular whether the super-horizon Majorana two-mode state can retain quantumness.

The paper is organized as follows. In Sec.~\ref{AxionInflationFermionModel} we introduce the axion-inflation background and derive the Majorana action on a torsion-free FLRW spacetime. In Sec.~\ref{QuantifyMajoranField} we quantize the Majorana field and show the equivalence between Hamiltonian diagonalization and the fermionic squeezing formalism. In Sec.~\ref{QuanInforMeasure} we construct the Majorana two-mode density matrix and evaluate the von Neumann entropy and logarithmic negativity as diagnostics of bipartite quantumness. We summarize the main results and discuss possible extensions in Sec.~\ref{ConcluDiscuss}. Technical details on spinor conventions, mode functions, Hamiltonian diagonalization, and the construction of the fermionic two-mode squeezed state are collected in the appendices.

\section{Majorana Fermion on a Torsion-Free FLRW Background \label{AxionInflationFermionModel}}

For simplicity, and as a toy-model setup, we consider the production of a Majorana-type fermion in an axion-monodromy inflationary background minimally coupled to Einstein gravity, namely 
\begin{align}
S_{\text{(GR-Axion)}}&=\!\int d^{4}x\sqrt{-g}\big\{\frac{M_{\text{pl}}^{2}}{2}R-\frac{1}{2}(\partial\phi)^{2}\!-\!V(\phi)\big\}\\
\nonumber
V(\phi)&=u^{3}(\sqrt{\phi^{2}+\phi_{c}^{2}}-\phi_{c}).
\end{align}Meanwhile, the dynamics of the Majorana fermion sector is governed by the action \cite{Adshead:2015kza,Freedman:2012zz}
\begin{small}
\begin{align}
\nonumber
S_{\text{(Majo)}}\!&=\!\int d^{4}x\sqrt{-g}\,\big\{-\tilde{\chi}_{\dot{\mathrm{I}}_{1}}^{\dagger}e_{~\tilde{a}}^{\mu}(\bar{\sigma}^{\tilde{a}})^{\dot{\mathrm{I}}_{1}\mathrm{J}_{1}}(\mathcal{D}_{\mu}\tilde{\chi})_{\mathrm{J}_{1}}+\frac{m}{2}\tilde{\chi}^{\mathrm{I}_{1}}\tilde{\chi}_{\mathrm{I}_{1}}\\
\label{MajoranaAction}
&+\frac{m}{2}\tilde{\chi}_{\dot{\mathrm{I}}_{1}}^{\dagger}\tilde{\chi}^{\dagger,\dot{\mathrm{I}}_{1}}+\frac{C_{\text{A-F}}}{f}\partial_{\mu}\phi\tilde{\chi}_{\dot{\mathrm{I}}_{1}}^{\dagger}e_{~\tilde{a}}^{\mu}(\bar{\sigma}^{\tilde{a}})^{\dot{\mathrm{I}}_{1}\mathrm{J}_{1}}\tilde{\chi}_{\mathrm{J}_{1}}\big\},
\end{align}
\end{small}in which $\mathrm{I},\mathrm{J},\dot{\mathrm{I}},\dot{\mathrm{J}}$ denote spinor indices in the Weyl (chiral) representation \cite{Dreiner:2008tw}, the Greek indices $\mu,\nu$ label spacetime components in the curved background, whereas the Latin indices with tildes, $\tilde{a},\tilde{b}$, refer to components in the local Lorentz frame with metric signature $\eta_{\tilde{a}\tilde{b}}=\text{Diag}(-,+,+,+)$. In addition, since we will make use of the two–component spinor conventions and techniques developed in \cite{Dreiner:2008tw}, it is convenient to introduce another set of Latin indices without tildes, $a,b$, to denote components in a local Lorentz frame with signature $\eta_{ab}=\text{Diag}(+,-,-,-)$ (the convention adopted in \cite{Dreiner:2008tw}). These two conventions are related through $\eta_{\tilde{a}\tilde{b}}=-\eta_{ab}$ and the corresponding relations for the sigma matrices, $\sigma^{\tilde{a}}=-\text{i}\sigma^{a},\sigma_{\tilde{a}}=\text{i}\sigma_{a}$ (and analogously for $\bar{\sigma}$). Moreover, the covariant derivatives acting on the spinor fields take the form
\begin{small}
\begin{align}
&(\mathcal{D}_{\mu})_{\mathrm{I}_{1}}^{~~\mathrm{I}_{2}}\!=\!\delta_{\mathrm{I}_{1}}^{\mathrm{I}_{2}}\partial_{\mu}\!-\frac{\text{i}}{2}\omega_{\mu\tilde{a}\tilde{b}}(\sigma^{\tilde{a}\tilde{b}})_{\mathrm{I}_{1}}^{~~\mathrm{I}_{2}},\\
&(\overleftarrow{\mathcal{D}}_{\mu}^{\dagger})_{~~\dot{\mathrm{I}}_{2}}^{\dot{\mathrm{I}}_{1}}\!\!=\!\delta_{\dot{\mathrm{I}}_{2}}^{\dot{\mathrm{I}}_{1}}\overleftarrow{\partial}_{\mu}\!+\frac{\text{i}}{2}\omega_{\mu\tilde{a}\tilde{b}}(\bar{\sigma}^{\tilde{a}\tilde{b}})_{~~\dot{\mathrm{I}}_{2}}^{\dot{\mathrm{I}}_{1}},
\end{align}
with 
\begin{align}
\sigma^{\tilde{a}\tilde{b}}\!=\!\frac{\text{i}}{4}(\sigma^{\tilde{a}}\bar{\sigma}^{\tilde{b}}-\sigma^{\tilde{b}}\bar{\sigma}^{\tilde{a}}),\bar{\sigma}^{\tilde{a}\tilde{b}}\!=\!\frac{\text{i}}{4}(\bar{\sigma}^{\tilde{a}}\sigma^{\tilde{b}}\!\!-\!\!\bar{\sigma}^{\tilde{b}}\sigma^{\tilde{a}}). \nonumber
\end{align}

\end{small}The torsion-free spin connection is constructed from the vierbein field $e_{~\,\tilde{a}}^{\mu}$ according to
\begin{small}
\begin{align}
\nonumber
\omega_{\mu\tilde{a}\tilde{b}}(e)&=\big(\eta_{[\tilde{a}\tilde{c}}\Gamma_{\mu\nu}^{\rho}e_{\rho}^{~\tilde{c}}e_{~\tilde{b}]}^{\nu}-\eta_{[\tilde{a}\tilde{c}}e_{~\tilde{b}]}^{\nu}\partial_{\mu}e_{\nu}^{~\tilde{c}}\big)\\
\nonumber
&=\frac{1}{2}e_{~\tilde{a}}^{\beta}(\partial_{\mu}e_{\beta\tilde{b}}-\partial_{\beta}e_{\mu\tilde{b}})-\frac{1}{2}e_{~\tilde{b}}^{\beta}(\partial_{\mu}e_{\beta\tilde{a}}-\partial_{\beta}e_{\mu\tilde{a}})\\
&-\frac{1}{2}e_{~\tilde{a}}^{\beta}e_{~\tilde{b}}^{\alpha}e_{\mu\tilde{c}}(\partial_{\beta}e_{\alpha}^{~\tilde{c}}-\partial_{\alpha}e_{\beta}^{~\tilde{c}}),
\end{align}
\end{small}in which the quantity $\Gamma^\rho_{\mu\nu}$ appearing in the first line denotes the Christoffel connection associated with the spacetime metric. The detailed derivation connecting the first line to the second line is presented explicitly in \cite{Freedman:2012zz,Lee:2025lyk}. In this work, we focus on an FLRW background and adopt conformal time coordinates defined by $d\tau = \frac{dt}{a}$, such that the spacetime metric takes the form
\begin{align}
\label{FLRWConforTimeMetric}
&ds^{2}=a(\tau)^{2}(-d\tau^{2}+d\boldsymbol{x}^{2})=e_{\mu}^{~\tilde{a}}e_{\nu}^{~\tilde{b}}\eta_{\tilde{a}\tilde{b}}dx^{\mu}dx^{\nu}.
\end{align}From \eqref{FLRWConforTimeMetric}, the vierbein field can be extracted as
\begin{align}
\label{VierbeinFLRW}
&e_{\mu}^{~\,\tilde{a}}=\left(\begin{array}{cccc}
a(\tau) & 0 & 0 & 0\\
0 & a(\tau) & 0 & 0\\
0 & 0 & a(\tau) & 0\\
0 & 0 & 0 & a(\tau)
\end{array}\right),\\
\label{InverseVierbeinFLRW}
&e_{~\,\tilde{a}}^{\mu}=\left(\begin{array}{cccc}
a(\tau)^{-1} & 0 & 0 & 0\\
0 & a(\tau)^{-1} & 0 & 0\\
0 & 0 & a(\tau)^{-1} & 0\\
0 & 0 & 0 & a(\tau)^{-1}
\end{array}\right).
\end{align}Note that it allows us to consistently employ the quasi–de Sitter approximation during the inflationary era, namely $a(\tau) = -\frac{1}{H\tau}$. Moreover, the action \eqref{MajoranaAction} contains terms involving the spin connection, which can be drastically simplified by making use of identities that relate the spin connection to the sigma matrices $\sigma^{ab}$ and $\bar{\sigma}^{ab}$. These identities are summarized in Appendix~\ref{SpinorInFlatSpacetime}, in particular the following one:
\begin{small}
\begin{align}
\label{ReduceSpinConnec}
\nonumber
\hspace{-1mm}(e_{~\tilde{a}}^{\mu}\bar{\sigma}^{\tilde{a}})^{\dot{\mathrm{I}}_{1}\mathrm{J}_{1}}(\omega_{\mu\tilde{a}_{1}\tilde{b}_{1}}\sigma^{\tilde{a}_{1}\tilde{b}_{1}})_{\mathrm{J}_{1}}^{~~\mathrm{J}_{2}}\! 
&=\!\frac{3a^{\prime}}{a^{2}}(\bar{\sigma}^{0})^{\dot{\mathrm{I}}_{1}\mathrm{J}_{2}}\! \\ 
&=\!\frac{3\text{i}a^{\prime}}{a^{2}}(\bar{\sigma}^{\tilde{0}})^{\dot{\mathrm{I}}_{1}\mathrm{J}_{2}}.
\end{align}
\end{small}After substituting the identity \eqref{ReduceSpinConnec} into the action \eqref{MajoranaAction} and performing the conformal field redefinition $\tilde{\chi}(\tau,\boldsymbol{x}) = a(\tau)^{-\frac{3}{2}}\chi(\tau,\boldsymbol{x})$, the spin-connection contribution is removed from the kinetic term.  The Majorana sector is then written in terms of the rescaled field as
\begin{align}
\nonumber
S_{\text{(Majo)}}&=\int d^{4}xa(\tau)\big\{\!-\!\chi_{\dot{\mathrm{I}}_{1}}^{\dagger}(e_{~\tilde{a}}^{\mu}\bar{\sigma}^{\tilde{a}})^{\dot{\mathrm{I}}_{1}\mathrm{J}_{1}}\partial_{\mu}\chi_{\mathrm{J}_{1}}\!-\frac{m}{2}\chi^{\mathrm{I}_{1}}\chi_{\mathrm{I}_{1}} \\
\label{MajoranaActionConforTime}
&-\!\frac{m}{2}\chi_{\dot{\mathrm{I}}_{1}}^{\dagger}\chi^{\dagger\dot{\mathrm{I}}_{1}}\!+\frac{\text{i}C_{\text{A-F}}}{f}\partial_{\mu}\bar{\phi}\chi_{\dot{\mathrm{I}}_{1}}^{\dagger}\!(e_{~\tilde{a}}^{\mu}\bar{\sigma}^{\tilde{a}})^{\dot{\mathrm{I}}_{1}\mathrm{J}_{1}}\chi_{\mathrm{J}_{1}}\!\big\}.
\end{align}This form makes the time dependence of the mass term and of the axial coupling to the rolling background explicit, and it will be the starting point for the mode-function analysis and canonical quantization below.  The equations of motion and the Fourier mode expansions of $\chi_{\mathrm{I}}(\tau,\boldsymbol{x})$, with the Bunch--Davies vacuum imposed as the initial condition, are derived explicitly in Appendix~\ref{AppendixB}.

\section{Canonical quantization and two equivalent formalisms \label{QuantifyMajoranField}}

To connect the mode functions derived above with particle production and, later, with quantum-information observables, we first construct the quadratic Hamiltonian in the operator basis associated with the Bunch--Davies vacuum at $\tau_0$.  It is useful to write the action \eqref{MajoranaActionConforTime} in a symmetrized form with the spinor indices displayed explicitly:
\begin{align}
\nonumber
S_{\text{(Majo)}}&=\!\int d^{4}x\big\{\frac{\text{i}}{2}\chi_{\dot{\mathrm{I}}_{1}}^{\dagger}(\bar{\sigma}^{0})^{\dot{\mathrm{I}}_{1}\mathrm{J}_{1}}\chi_{\mathrm{J}_{1}}^{\prime}+\frac{\text{i}}{2}\chi^{\mathrm{J}_{1}}(\sigma^{0})_{\mathrm{J}_{1}\dot{\mathrm{I}}_{1}}\chi^{\prime\dagger\dot{\mathrm{I}}_{1}}\\
\nonumber
&+\frac{C_{\text{A-F}}\bar{\phi}^{\prime}}{2f}\big(\chi_{\dot{\mathrm{I}}_{1}}^{\dagger}(\bar{\sigma}^{0})^{\dot{\mathrm{I}}_{1}\mathrm{J}_{1}}\chi_{\mathrm{J}_{1}}\!-\!\chi^{\mathrm{J}_{1}}(\sigma^{0})_{\mathrm{J}_{1}\dot{\mathrm{I}}_{1}}\chi^{\dagger\dot{\mathrm{I}}_{1}}\big)\\
\nonumber
&+\frac{\text{i}}{2}\big(\chi_{\dot{\mathrm{I}}_{1}}^{\dagger}(\bar{\sigma}^{i})^{\dot{\mathrm{I}}_{1}\mathrm{J}_{1}}\partial_{i}\chi_{\mathrm{J}_{1}}+\chi^{\mathrm{J}_{1}}(\sigma^{i})_{\mathrm{J}_{1}\dot{\mathrm{I}}_{1}}\partial_{i}\chi^{\dagger\dot{\mathrm{I}}_{1}}\big)\\
\label{MajoranaActionNormal}
&-\frac{a(\tau)m}{2}(\chi^{\mathrm{I}_{1}}\chi_{\mathrm{I}_{1}}+\chi_{\dot{\mathrm{I}}_{1}}^{\dagger}\chi^{\dagger\dot{\mathrm{I}}_{1}})\big\}.
\end{align}In deriving \eqref{MajoranaActionNormal}, we used the transpose identity
\begin{align}
&\chi_{\dot{\mathrm{I}}_{1}}^{\dagger}(x)(\bar{\sigma}^{\mu})^{\dot{\mathrm{I}}_{1}\mathrm{J}_{1}}\xi_{\mathrm{J}_{1}}(x)=-\xi^{\mathrm{J}_{1}}(x)(\sigma^{\mu})_{\mathrm{J}_{1}\dot{\mathrm{I}}_{1}}\chi^{\dagger\dot{\mathrm{I}}_{1}}(x),
\end{align}together with integrations by parts. The conjugate momentum is therefore
\begin{align}
&\pi_{(\chi)}^{\mathrm{J}_{1}}(x)=S_{\text{(Majo)}}\frac{\overleftarrow{\delta}}{\overleftarrow{\delta}\chi_{\mathrm{J}_{1}}^{\prime}(x)}=\text{i}\chi_{\dot{\mathrm{I}}_{1}}^{\dagger}(x)(\bar{\sigma}^{0})^{\dot{\mathrm{I}}_{1}\mathrm{J}_{1}},
\end{align}and the Hamiltonian density is defined by
\begin{align}
\label{DefineMajoranaDensity}
&\mathcal{H}_{\text{(Majo)}}=\frac{1}{2}\pi_{(\chi)}^{\mathrm{I}_{1}}\chi_{\mathrm{I}_{1}}^{\prime}+\frac{1}{2}\chi_{\dot{\mathrm{I}}_{1}}^{\prime\dagger}\pi_{(\chi)}^{\dot{\mathrm{I}}_{1}}-\mathcal{L}_{\text{(Majo)}}.
\end{align}We then promote $\chi_{\mathrm{I}_{1}}(\tau,\boldsymbol{x})$ to the field operator $\hat{\chi}_{\mathrm{I}_{1}}(\tau,\boldsymbol{x})$ and use the Fourier expansion
\begin{align}
\nonumber
\hat{\chi}_{\mathrm{I}_{1}}(\tau,\boldsymbol{x})&=\!\!\!\sum_{s=\pm\frac{1}{2}}\!\!\int\!\frac{d^{3}\boldsymbol{k}}{(2\pi)^{3}}\big(x_{\mathrm{I}_{1},\boldsymbol{k}}^{(s)}(\tau)\text{e}^{\text{i}\boldsymbol{k}\cdot\boldsymbol{x}}\hat{a}_{\boldsymbol{k}}^{(s)}(\tau_{0})\\
\label{SpinorChiModeExpan}
&+y_{\mathrm{I}_{1},\boldsymbol{k}}^{(s)}(\tau)\text{e}^{-\text{i}\boldsymbol{k}\cdot\boldsymbol{x}}\hat{a}_{\boldsymbol{k}}^{(s)\dagger}(\tau_{0})\big).
\end{align}The operators carrying the time label $\tau_0$ annihilate or create excitations with respect to the Bunch--Davies vacuum defined in the asymptotic past. Using the helicity eigenspinors summarized in Appendix~\ref{SpinorInFlatSpacetime}, the spinor mode functions are further decomposed as
\begin{align}
\nonumber
&x_{\dot{\mathrm{I}}_{1},\boldsymbol{k}}^{(s)\dagger}(\tau)\!\!=\!\!X_{k}^{(s)\star}(\tau)\xi_{s,\dot{\mathrm{I}}_{1}}^{\dagger}(\tilde{\boldsymbol{k}}),y_{\boldsymbol{k}}^{\dagger(s),\dot{\mathrm{I}}_{1}}(\tau)\!\!=\!\!Y_{k}^{\star(s)}(\tau)\xi_{s}^{\dot{\mathrm{I}}_{1}}(\tilde{\boldsymbol{k}}),\\
\label{EigenSpinorDecompose}
&x_{\mathrm{I}_{1},\boldsymbol{k}}^{(s)}(\tau)\!\!=\!\!X_{k}^{(s)}(\tau)\xi_{s,\mathrm{I}_{1}}(\tilde{\boldsymbol{k}}),y_{\boldsymbol{k}}^{(s),\mathrm{I}_{1}}(\tau)\!\!=\!\!Y_{k}^{(s)}(\tau)\xi_{s}^{\dagger\mathrm{I}_{1}}(\tilde{\boldsymbol{k}}).
\end{align}The scalar functions $X_k^{(s)}$ and $Y_k^{(s)}$ obey the Wronskian normalization \eqref{NormalEigenSpinor}, which follows from the equal-time canonical anti-commutator.  Substituting \eqref{SpinorChiModeExpan} and \eqref{EigenSpinorDecompose} into \eqref{DefineMajoranaDensity} gives the quadratic Hamiltonian operator \begin{align}
\nonumber
\hat{H}_{\text{(Majo)}}\!(\tau)&=\!\!\sum_{s=\pm\frac{1}{2}}\!\int_{\mathbb{R}^{3+}}\!\!\frac{d^{3}\boldsymbol{k}}{(2\pi)^{3}}\big\{\mathcal{A}_{k}^{(s)}(\tau)\hat{a}_{\boldsymbol{k}}^{(s)\dagger}(\tau_{0})\hat{a}_{\boldsymbol{k}}^{(s)}(\tau_{0})\\
\nonumber
&-\!\mathcal{A}_{k}^{(s)}(\tau)\hat{a}_{-\boldsymbol{k}}^{(s)}(\tau_{0})\hat{a}_{-\boldsymbol{k}}^{(s)\dagger}(\tau_{0})\!+\!\mathcal{B}_{k}^{(s)}(\tau)\hat{a}_{-\boldsymbol{k}}^{(s)}(\tau_{0})\hat{a}_{\boldsymbol{k}}^{(s)}(\tau_{0})\\
&+\mathcal{B}_{k}^{(s)\star}(\tau)\hat{a}_{\boldsymbol{k}}^{(s)\dagger}(\tau_{0})\hat{a}_{-\boldsymbol{k}}^{(s)\dagger}(\tau_{0})\big\}.
\end{align}Here $\mathbb{R}^{3+}$ denotes one half of momentum space, so that each pair $(\boldsymbol{k},-\boldsymbol{k})$ is counted once.  The real coefficient $\mathcal{A}_{k}^{(s)}(\tau)$ and the generally complex coefficient $\mathcal{B}_{k}^{(s)}(\tau)$ are determined by $X_k^{(s)}$ and $Y_k^{(s)}$; their explicit expressions are given in Appendix~\ref{AppendixC}, see Eq.~\eqref{QuarticHamiltonMatrix}.  The terms proportional to $\mathcal B_k^{(s)}$ and $\mathcal B_k^{(s)\star}$ are the pair-annihilation and pair-creation terms, and therefore encode the nonadiabatic production of correlated Majorana pairs.

\subsection{Hamiltonian diagonalization and the Bogoliubov transformation of creation and annihilation operators}

More technical details of the Hamiltonian construction, the explicit form of $\boldsymbol{\mathcal{M}}_{k}^{(s)}(\tau)$, and its diagonalization are given in Appendix~\ref{AppendixC}. The standard way to define instantaneous quasiparticles is to diagonalize the quadratic Hamiltonian for each momentum pair and helicity sector.  This is the fermionic version of the Bogoliubov transformation: a canonical linear recombination of creation and annihilation operators that preserves the anti-commutation relations and turns a paired quadratic Hamiltonian into a diagonal quasiparticle form.  In curved spacetime, the same idea measures the mismatch between the positive-frequency basis selected in the asymptotic Bunch--Davies region and the instantaneous basis at time $\tau$ \cite{Birrell:1982ix,Kanno:2016qcc}; for the axion-inflation fermion system, the corresponding Hamiltonian coefficients are built from the Majorana mode functions discussed above \cite{Adshead:2015kza}.

In the Nambu basis $(\hat a_{\boldsymbol{k}}^{(s)},\hat a_{-\boldsymbol{k}}^{(s)\dagger})^{\mathrm T}$, the Hamiltonian is governed by the Hermitian traceless matrix
\begin{align}
&\boldsymbol{\mathcal{M}}_{k}^{(s)}(\tau)=\left(\begin{array}{cc}
\mathcal{A}_{k}^{(s)}(\tau) & \mathcal{B}_{k}^{(s)\star}(\tau)\\
\mathcal{B}_{k}^{(s)}(\tau) & -\mathcal{A}_{k}^{(s)}(\tau)
\end{array}\right).
\end{align}
Its eigenvalues are $\pm \omega_k^{(s)}(\tau)$, with
\begin{align}
\label{OmegaRelateToAB}
\nonumber
(\omega_{k}^{(s)})^{2}\!&=\!(\mathcal{A}_{k}^{(s)})^{2}+\vert\mathcal{B}_{k}^{(s)}\vert^{2}\!\\
&=\!(sk\!+\!\frac{C_{\text{A-F}}\bar{\phi}^{\prime}}{2f})^{2}\!+\!\frac{m^{2}a^{2}}{4}.
\end{align}
The diagonalization is implemented by a unitary matrix $\boldsymbol{\mathcal U}_{k}^{(s)}(\tau)$ satisfying
\begin{align}
&\boldsymbol{\mathcal{M}}_{k}^{(s)}(\tau)=
\boldsymbol{\mathcal{U}}_{k}^{(s)\dagger}(\tau)
\left(\begin{array}{cc}
\omega_{k}^{(s)}(\tau) & 0\\
0 & -\omega_{k}^{(s)}(\tau)
\end{array}\right)
\boldsymbol{\mathcal{U}}_{k}^{(s)}(\tau),\\
&\boldsymbol{\mathcal{U}}_{k}^{(s)}(\tau)=\frac{1}{\sqrt{2\omega_{k}^{(s)}}}\left(\begin{array}{cc}
\frac{\mathcal{B}_{k}^{(s)}}{\sqrt{\omega_{k}^{(s)}-\mathcal{A}_{k}^{(s)}}} & \sqrt{\omega_{k}^{(s)}-\mathcal{A}_{k}^{(s)}}\\
\frac{\mathcal{B}_{k}^{(s)}}{\sqrt{\omega_{k}^{(s)}+\mathcal{A}_{k}^{(s)}}} & -\sqrt{\omega_{k}^{(s)}+\mathcal{A}_{k}^{(s)}}
\end{array}\right).
\end{align}
More details of this diagonalization, including the explicit expressions for $\mathcal A_k^{(s)}$ and $\mathcal B_k^{(s)}$, are collected in Appendix~\ref{AppendixC}.  Acting with $\boldsymbol{\mathcal U}_{k}^{(s)}$ on the Nambu basis defines the time-dependent annihilation and creation operators,
\begin{align}
\left(\begin{array}{c}
\hat{a}_{\boldsymbol{k}}^{(s)}(\tau)\\
\hat{a}_{-\boldsymbol{k}}^{(s)\dagger}(\tau)
\end{array}\right)
=\boldsymbol{\mathcal{U}}_{k}^{(s)}(\tau)
\left(\begin{array}{c}
\hat{a}_{\boldsymbol{k}}^{(s)}(\tau_{0})\\
\hat{a}_{-\boldsymbol{k}}^{(s)\dagger}(\tau_{0})
\end{array}\right).
\end{align}
Equivalently, this gives the Bogoliubov transformation
\begin{align}
\nonumber
\hat{a}_{\boldsymbol{k}}^{(s)}(\tau)&=\frac{\mathcal{B}_{k}^{(s)}}{\sqrt{2\omega_{k}^{(s)}}\sqrt{\omega_{k}^{(s)}-\mathcal{A}_{k}^{(s)}}}\hat{a}_{\boldsymbol{k}}^{(s)}(\tau_{0})\\
\label{StandAnnihilaBogoliubov}
&+\frac{\sqrt{\omega_{k}^{(s)}-\mathcal{A}_{k}^{(s)}}}{\sqrt{2\omega_{k}^{(s)}}}\hat{a}_{-\boldsymbol{k}}^{(s)\dagger}(\tau_{0}),\\
\nonumber
\hat{a}_{-\boldsymbol{k}}^{(s)\dagger}(\tau)&=\frac{\mathcal{B}_{k}^{(s)}}{\sqrt{2\omega_{k}^{(s)}}\sqrt{\omega_{k}^{(s)}+\mathcal{A}_{k}^{(s)}}}\hat{a}_{\boldsymbol{k}}^{(s)}(\tau_{0}),\\
\label{StandCreationBogoliubov}
&-\frac{\sqrt{\omega_{k}^{(s)}+\mathcal{A}_{k}^{(s)}}}{\sqrt{2\omega_{k}^{(s)}}}\hat{a}_{-\boldsymbol{k}}^{(s)\dagger}(\tau_{0}).
\end{align}
This transformation preserves the canonical anti-commutation relations.  Its off-diagonal entries quantify the mixing of $\hat a_{\boldsymbol{k}}^{(s)}(\tau_0)$ with $\hat a_{-\boldsymbol{k}}^{(s)\dagger}(\tau_0)$, and hence the production of Majorana pairs from the initial Bunch--Davies vacuum.  The particle number in the instantaneous basis is therefore
\begin{align}
\nonumber
\mathcal{N}_{k}^{(s)}(\tau)
&=\,_{\tau_0}\langle0|\hat a_{\boldsymbol{k}}^{(s)\dagger}(\tau)\hat a_{\boldsymbol{k}}^{(s)}(\tau)|0\rangle_{\tau_0}\\
&=\frac{\omega_k^{(s)}(\tau)-\mathcal A_k^{(s)}(\tau)}{2\omega_k^{(s)}(\tau)}.
\end{align}
Thus $\mathcal B_k^{(s)}=0$ corresponds to an adiabatic, particle-free basis, while a nonzero $\mathcal B_k^{(s)}$ measures pair mixing induced by the time-dependent background.  The same result will now be recast in a squeezing language, which is more directly adapted to the two-mode state used in Sec.~\ref{QuanInforMeasure}.

\subsection{Fermionic squeezing formalism}

The Hamiltonian above can also be organized by the $su(2)$ algebra generated by pair creation, pair annihilation, and number rotation.  Fermionic squeezed, paired, or BCS-like two-mode structures are well established in condensed-matter applications, most notably in the BCS description of superconducting pairing, and have also been used in fermionic Gaussian-state and quantum-information settings \cite{Bardeen:1957mv,Botero:2004}.  Here we adapt the same algebraic logic as an operator-level bridge between primordial Majorana production and the two-mode entanglement measures.  The detailed construction is given in Appendix~\ref{SqueeTwoModeBuild}.  For each $(\boldsymbol{k},-\boldsymbol{k})$ pair, we define
\begin{align}
&\hat{\mathcal{J}}_{+,\boldsymbol{k}}^{(s)}=\hat{a}_{-\boldsymbol{k}}^{(s)\dagger}(\tau_{0})\hat{a}_{\boldsymbol{k}}^{(s)\dagger}(\tau_{0}),~
\hat{\mathcal{J}}_{-,\boldsymbol{k}}^{(s)}=\hat{a}_{\boldsymbol{k}}^{(s)}(\tau_{0})\hat{a}_{-\boldsymbol{k}}^{(s)}(\tau_{0}),\\
&\hat{\mathcal{J}}_{z,\boldsymbol{k}}^{(s)}=\frac{1}{2}\big(\hat{a}_{-\boldsymbol{k}}^{(s)\dagger}(\tau_{0})\hat{a}_{-\boldsymbol{k}}^{(s)}(\tau_{0})-\hat{a}_{\boldsymbol{k}}^{(s)}(\tau_{0})\hat{a}_{\boldsymbol{k}}^{(s)\dagger}(\tau_{0})\big),
\end{align}
which satisfy the standard $su(2)$ commutation relations.  In this basis, the quadratic Hamiltonian takes the compact form
\begin{align}
\nonumber
\hat{H}_{\text{(Majo)}}(\tau)=&
\sum_{s=\pm\frac{1}{2}}\int_{\mathbb{R}^{3+}}\frac{d^{3}\boldsymbol{k}}{(2\pi)^{3}}\big\{-\mathcal{B}_{k}^{(s)\star}\hat{\mathcal{J}}_{+,\boldsymbol{k}}^{(s)}\\
&-\mathcal{B}_{k}^{(s)}\hat{\mathcal{J}}_{-,\boldsymbol{k}}^{(s)}+2\mathcal{A}_{k}^{(s)}\hat{\mathcal{J}}_{z,\boldsymbol{k}}^{(s)}\big\}.
\end{align}
Thus the time-evolution operator for each pair is an $SU(2)$ element.  It may be parameterized as a fermionic squeezing operator followed by a number rotation,
\begin{align}
&\hat{\mathcal{U}}(\tau,\tau_{0})=\hat{\mathcal{S}}(r,\varphi)\hat{\mathcal{R}}(\omega),\\ \nonumber
&\hat{\mathcal{S}}(r,\varphi)=\exp\!\bigg[\text{i}\sum_s\int_{\mathbb{R}^{3+}}\frac{d^3\boldsymbol{k}}{(2\pi)^3}r_k^{(s)}(\tau)\\
&\hspace{1.2cm}\times\big(\text{e}^{-\text{i}\varphi_k^{(s)}(\tau)}\hat{\mathcal{J}}_{+,\boldsymbol{k}}^{(s)}+\text{e}^{\text{i}\varphi_k^{(s)}(\tau)}\hat{\mathcal{J}}_{-,\boldsymbol{k}}^{(s)}\big)\bigg],\\
&\hat{\mathcal{R}}(\omega)=\exp\!\bigg[2\text{i}\sum_s\int_{\mathbb{R}^{3+}}\frac{d^3\boldsymbol{k}}{(2\pi)^3}\omega_k^{(s)}(\tau)\hat{\mathcal{J}}_{z,\boldsymbol{k}}^{(s)}\bigg].
\end{align}
The parameters $r_k^{(s)}$, $\varphi_k^{(s)}$, and $\omega_k^{(s)}$ are fixed by matching the operator transformation generated by $\hat{\mathcal U}$ to the Hamiltonian-diagonalization result above.  Concretely, the time-dependent operators are obtained from the initial operators through
\begin{align}
\nonumber
\hat a_{\boldsymbol{k}}^{(s)}(\tau)
&=\hat{\mathcal U}^{-1}(\tau,\tau_0)\hat a_{\boldsymbol{k}}^{(s)}(\tau_0)\hat{\mathcal U}(\tau,\tau_0)\\ 
&=\hat{\mathcal R}^{-1}(\omega)\hat{\mathcal S}^{-1}(r,\varphi)\hat a_{\boldsymbol{k}}^{(s)}(\tau_0)\hat{\mathcal S}(r,\varphi)\hat{\mathcal R}(\omega).
\end{align}
Using the Baker--Campbell--Hausdorff expansion together with the $su(2)$ commutators, the squeezing part first gives
\begin{align}
\nonumber
\hat{\mathcal S}^{-1}\hat a_{\boldsymbol{k}}^{(s)}(\tau_0)\hat{\mathcal S}
=&\cos r_k^{(s)}\,\hat a_{\boldsymbol{k}}^{(s)}(\tau_0)\\
&
-\text{i}\text{e}^{-\text{i}\varphi_k^{(s)}}\sin r_k^{(s)}\,\hat a_{-\boldsymbol{k}}^{(s)\dagger}(\tau_0),
\end{align}
while the number rotation contributes the phase factors
\begin{align}
&\hat{\mathcal R}^{-1}\hat a_{\boldsymbol{k}}^{(s)}(\tau_0)\hat{\mathcal R}
=\text{e}^{\text{i}\omega_k^{(s)}}\hat a_{\boldsymbol{k}}^{(s)}(\tau_0),\\
&\hat{\mathcal R}^{-1}\hat a_{-\boldsymbol{k}}^{(s)\dagger}(\tau_0)\hat{\mathcal R}
=\text{e}^{-\text{i}\omega_k^{(s)}}\hat a_{-\boldsymbol{k}}^{(s)\dagger}(\tau_0).
\end{align}
The normal-ordering and disentangling steps behind these identities are shown explicitly in Appendix~\ref{SqueeTwoModeBuild}.  Combining the two steps gives
\begin{align}
\nonumber
\hat{a}_{\boldsymbol{k}}^{(s)}(\tau)&=\cos(r_{k}^{(s)}(\tau))\text{e}^{\text{i}\omega_{k}^{(s)}(\tau)}\hat{a}_{\boldsymbol{k}}^{(s)}(\tau_{0})\\
\label{SqueezedAnnihila}
&-\text{i}\sin(r_{k}^{(s)}(\tau))\text{e}^{-\text{i}\big(\omega_{k}^{(s)}(\tau)+\varphi_{k}^{(s)}(\tau)\big)}\hat{a}_{-\boldsymbol{k}}^{(s)\dagger}(\tau_{0}),\\
\nonumber
\hat{a}_{-\boldsymbol{k}}^{(s)\dagger}(\tau)&=\cos(r_{k}^{(s)}(\tau))\text{e}^{-\text{i}\omega_{k}^{(s)}(\tau)}\hat{a}_{-\boldsymbol{k}}^{(s)\dagger}(\tau_{0})\\
\label{SqueezedCreation}
&+\text{i}\sin(r_{k}^{(s)}(\tau))\text{e}^{\text{i}\big(\omega_{k}^{(s)}(\tau)+\varphi_{k}^{(s)}(\tau)\big)}\hat{a}_{\boldsymbol{k}}^{(s)}(\tau_{0}).
\end{align}
Acting on the Bunch--Davies vacuum, the same evolution produces a fermionic two-mode squeezed state,
\begin{align}
\label{FermiTwoModeState}
\vert\Psi_{\boldsymbol{k}}^{(s)}(\tau)\rangle&=\cos\big(r_{k}^{(s)}(\tau)\big)\text{e}^{-\text{i}\omega_{k}^{(s)}(\tau)}\vert0_{\boldsymbol{k}}^{(s)},0_{-\boldsymbol{k}}^{(s)}\rangle_{\tau_{0}}\nonumber\\
&\hspace{-0.8cm}+\text{i}\sin\big(r_{k}^{(s)}(\tau)\big)\text{e}^{-\text{i}\big(\varphi_{k}^{(s)}(\tau)+\omega_{k}^{(s)}(\tau)\big)}\vert1_{\boldsymbol{k}}^{(s)},1_{-\boldsymbol{k}}^{(s)}\rangle_{\tau_{0}}.
\end{align}
Because of the Pauli principle, this state contains only the vacuum and one-pair sectors.  This finite-dimensional structure is the main reason why the Majorana pair provides a clean platform for the quantum-information measures introduced in the next section.  Finally, matching \eqref{StandAnnihilaBogoliubov}--\eqref{StandCreationBogoliubov} with \eqref{SqueezedAnnihila}--\eqref{SqueezedCreation} yields
\begin{align}
\label{SqueezeAmplitude}
&\mathcal{N}_{k}^{(s)}(\tau)=\sin(r_{k}^{(s)}(\tau))^{2}=\frac{\omega_{k}^{(s)}(\tau)-\mathcal{A}_{k}^{(s)}(\tau)}{2\omega_{k}^{(s)}(\tau)},\\ \nonumber
\label{SqueezeRotationAngles}
&\tan(2\omega_{k}^{(s)})\!=\!-\tan(2\varphi_{k}^{(s)})\!\\ 
&\hspace{1.56cm}=\!\text{Im}\big((\mathcal{B}_{k}^{(s)})^{2}\big)\big/\text{Re}\big((\mathcal{B}_{k}^{(s)})^{2}\big).
\end{align}
Here $\mathcal{N}_{k}^{(s)}$ is the instantaneous occupation number of the Majorana mode, while $r_k^{(s)}$, $\varphi_k^{(s)}$, and $\omega_k^{(s)}$ determine the amplitude and phases of the two-mode state.

\section{Quantum-information diagnostics of primordial Majorana two-mode states \label{QuanInforMeasure}}

We now use the fermionic two-mode squeezed state derived in Sec.~\ref{QuantifyMajoranField} and Appendix~\ref{SqueeTwoModeBuild} to quantify the residual quantumness of primordial Majorana modes.  Quantum fields in expanding backgrounds have long been known to produce particles through the mismatch between early- and late-time notions of positive frequency \cite{Parker:1968mv,Birrell:1982ix}.  Fermionic fields provide a particularly sharp version of this problem: their excitations are Pauli-limited, their pair-production amplitudes are naturally described by Bogoliubov transformations, and their two-mode Hilbert space is finite.  This perspective has been used in studies of fermion production during preheating and axion inflation \cite{Greene:1998nh,Adshead:2015kza,Adshead:2018oaa}, as well as in analyses of Dirac-field entanglement, Bell-type violation, quantum discord, and electromagnetic-background effects in de Sitter space \cite{Kanno:2016qcc,Bhattacharya:2020bal,Ali:2021jch}.  In parallel, quantum-information tools have become useful diagnostics for the quantum-to-classical problem of inflationary perturbations, especially for tracking entanglement and decoherence beyond a purely occupation-number description \cite{Martin:2012pea,Brahma:2020zpk,Brahma:2024ycc}.  A closely related program asks whether primordial correlations can be certified through Bell-type or nonclassicality witnesses.  Early discussions of Bell inequalities for inflationary spectra and the quantum information carried by cosmological correlations led to Maldacena's explicit model of cosmological Bell inequalities, while later analyses clarified both the promise and the obstructions of such tests once realistic observables and decoherence effects are included \cite{Campo:2005sv,Lim:2014uea,Maldacena:2015bha,Martin:2015qta,Martin:2017zxs,Sou:2024tjv}.  Our goal here is to combine these two viewpoints: the Majorana mode functions determine the fermionic squeezing parameter, while the von Neumann entropy and logarithmic negativity diagnose the bipartite quantum correlations of the pair $(\boldsymbol{k},-\boldsymbol{k})$.  The logarithmic negativity should not be read as an observational Bell test by itself, but as a state-level measure of the fermionic inseparability that any later classicalization or decoherence mechanism must erase, suppress, or hide.

When the density matrix is written in matrix form, we adopt the following ordering convention for the occupation-number basis,
\begin{align}
\label{TwoModeBasisConvention}
    \left\{
    |0\rangle \otimes \langle 0 \vert,\,
    |0\rangle \otimes \langle 1 \vert,\,
    |1\rangle \otimes \langle 0 \vert,\,
    |1 \rangle \otimes \langle 1 \vert
    \right\},
\end{align}
the density matrix constructed from the fermionic two-mode state \eqref{FermiTwoModeState}, after summing over the helicity sectors, can be written as
\begin{align}
\nonumber
\hat{\rho}_{\boldsymbol{k},-\boldsymbol{k}}&=\sum_{s}\big(\cos(r_{k}^{(s)}(\tau))\text{e}^{-\text{i}\omega_{k}^{(s)}(\tau)}\vert0_{\boldsymbol{k}}^{(s)},0_{-\boldsymbol{k}}^{(s)}\rangle_{\tau_{0}}\\
\nonumber
&+\text{i}\sin(r_{k}^{(s)}(\tau))\text{e}^{-\text{i}\big(\varphi_{k}^{(s)}(\tau)+\omega_{k}^{(s)}(\tau)\big)}\vert1_{\boldsymbol{k}}^{(s)},1_{-\boldsymbol{k}}^{(s)}\rangle_{\tau_{0}}\big)\\
\nonumber
&\otimes\sum_{s^{\prime}}\big(\cos(r_{k}^{(s^{\prime})}(\tau))\text{e}^{\text{i}\omega_{k}^{(s^{\prime})}(\tau)}{}_{\tau_{0}\!}\langle0_{\boldsymbol{k}}^{(s^{\prime})},0_{-\boldsymbol{k}}^{(s^{\prime})}\vert\\
\label{DensityMatrixSSprime}
&\hspace{-3mm}-\text{i}\sin(r_{k}^{(s^{\prime})}(\tau))\text{e}^{\text{i}\big(\varphi_{k}^{(s^{\prime})}(\tau)+\omega_{k}^{(s^{\prime})}(\tau)\big)}{}_{\tau_{0}\!}\langle1_{\boldsymbol{k}}^{(s)},1_{-\boldsymbol{k}}^{(s)}\vert\big).
\end{align}
This density matrix is the starting point for the two complementary diagnostics below.  The von Neumann entropy measures the entanglement entropy of the reduced state, whereas the logarithmic negativity directly probes the failure of separability through the partial transpose.

\subsection{Von Neuman entropy}

\begin{figure}[H]
 	\begin{center}
 		\includegraphics[scale=0.64]{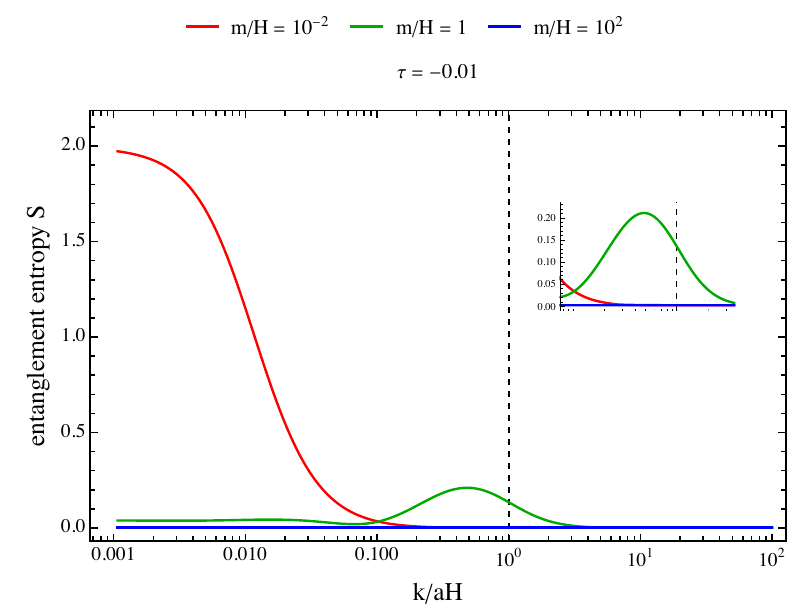}
        \caption{Von Neumann entropy \eqref{VonNeumanFermiTwoMode} of the reduced Majorana two-mode state near the end of inflation, evaluated at $\tau=-10^{-2}$ with $C_{\text{A-F}}=10^{-6}$, $f=10^{-3}$, $M_{\text{pl}}=1$, and $\epsilon=10^{-3}$.  The Hubble scale is fixed by the CMB normalization of the curvature power spectrum, $\mathcal{P}^{\star}_{\mathcal{R}}=H^2/\big(8\pi^2M_{\text{pl}}^2\epsilon\big)=2.1\times10^{-9}$, giving $H=M_{\text{pl}}\sqrt{8\pi^2\epsilon\times2.1\times10^{-9}}$.  The inset highlights the super-horizon region, where lighter Majorana modes retain a larger entanglement entropy.}
        \label{VonNeumanEntropyForPriMajorana}
 	\end{center}
 \end{figure}

We first trace over the $-\boldsymbol{k}$ mode in the full density matrix $\hat{\rho}_{\boldsymbol{k},-\boldsymbol{k}}$.  The reduced density matrix for the $\boldsymbol{k}$ subsystem is \begin{align}
\label{DefineHatRhoOnlyK}
\hat{\rho}_{\boldsymbol{k}}&=\sum_{\lambda=\pm\frac{1}{2}}\sum_{n=0,1}\langle n_{-\boldsymbol{k}}^{(\lambda)}\vert\hat{\rho}_{\boldsymbol{k},-\boldsymbol{k}}\vert n_{-\boldsymbol{k}}^{(\lambda)}\rangle\\
\nonumber
&=\cos^{2}(r_{k}^{(+\frac{1}{2})}(\tau))\vert0_{\boldsymbol{k}}^{(+\frac{1}{2})}\rangle\otimes\langle0_{\boldsymbol{k}}^{(+\frac{1}{2})}\vert\\
\nonumber
&+\cos^{2}(r_{k}^{(-\frac{1}{2})}(\tau))\vert0_{\boldsymbol{k}}^{(-\frac{1}{2})}\rangle\otimes\langle0_{\boldsymbol{k}}^{(-\frac{1}{2})}\vert\\
\nonumber
&+\sin^{2}(r_{k}^{(+\frac{1}{2})}(\tau))\vert1_{\boldsymbol{k}}^{(+\frac{1}{2})}\rangle\otimes\langle1_{\boldsymbol{k}}^{(+\frac{1}{2})}\vert\\
\nonumber
&+\sin^{2}(r_{k}^{(-\frac{1}{2})}(\tau))\vert1_{\boldsymbol{k}}^{(-\frac{1}{2})}\rangle\otimes{}_{\tau_{0}\!}\langle1_{\boldsymbol{k}}^{(-\frac{1}{2})}\vert.
\end{align}
In the basis convention \eqref{TwoModeBasisConvention}, this reduced density matrix is diagonal and can be viewed as the direct sum of the two helicity sectors.  The corresponding von Neumann entropy is therefore
\begin{align}
\nonumber
S[\hat{\rho}_{\boldsymbol{k}}]\!\!=&\!-\text{Tr}\big\{\hat{\rho}_{\boldsymbol{k}}\log_{2}(\hat{\rho}_{\boldsymbol{k}})\big\}\!\!=\!-(1\!\!-\!\mathcal{N}_{k}^{(+\frac{1}{2})})\log_{2}(1\!\!-\!\mathcal{N}_{k}^{(+\frac{1}{2})})\\
\nonumber
&-\!(1\!-\!\mathcal{N}_{k}^{(-\frac{1}{2})})\log_{2}(1\!-\!\mathcal{N}_{k}^{(-\frac{1}{2})})\!-\!\mathcal{N}_{k}^{(+\frac{1}{2})}\log_{2}(\mathcal{N}_{k}^{(+\frac{1}{2})})\\
\label{VonNeumanFermiTwoMode}
&-\mathcal{N}_{k}^{(-\frac{1}{2})}\log_{2}(\mathcal{N}_{k}^{(-\frac{1}{2})}),
\end{align}
where the particle number $\mathcal{N}_{k}^{(s)}$ is given in \eqref{SqueezeAmplitude}.  Since the state \eqref{FermiTwoModeState} is a bipartite two-mode state, the entropy of the reduced density matrix measures the entanglement between the $\boldsymbol{k}$ and $-\boldsymbol{k}$ sectors.  Figure~\ref{VonNeumanEntropyForPriMajorana} shows the dependence of $S[\hat{\rho}_{\boldsymbol{k}}]$ on the wave number near the end of inflation.  The light Majorana modes retain stronger super-horizon quantum memory, while modes with masses comparable to the Hubble scale are more sensitive to the nonadiabatic dynamics around horizon crossing.

\subsection{Logarithmic negativity}

\begin{figure}[ht]
 	\begin{center}
         \includegraphics[scale=0.64]{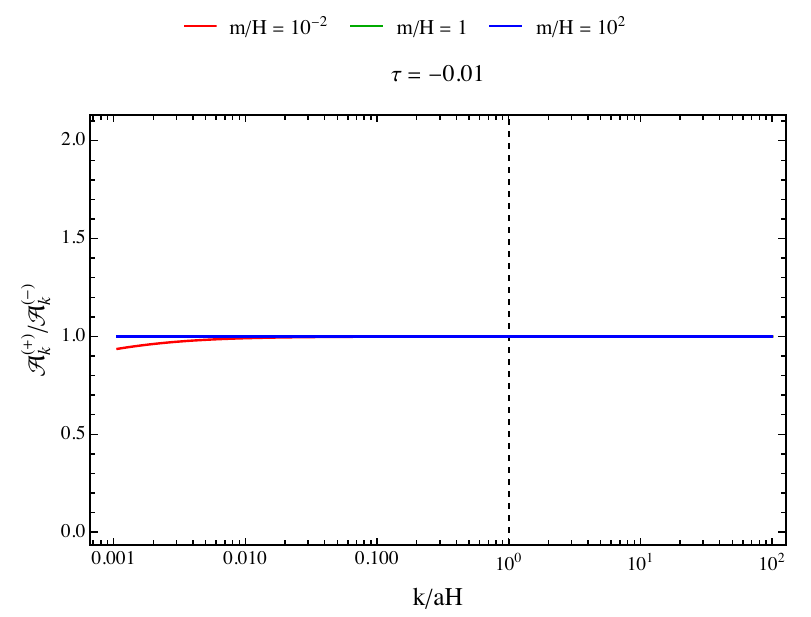}\\
         \includegraphics[scale=0.64]{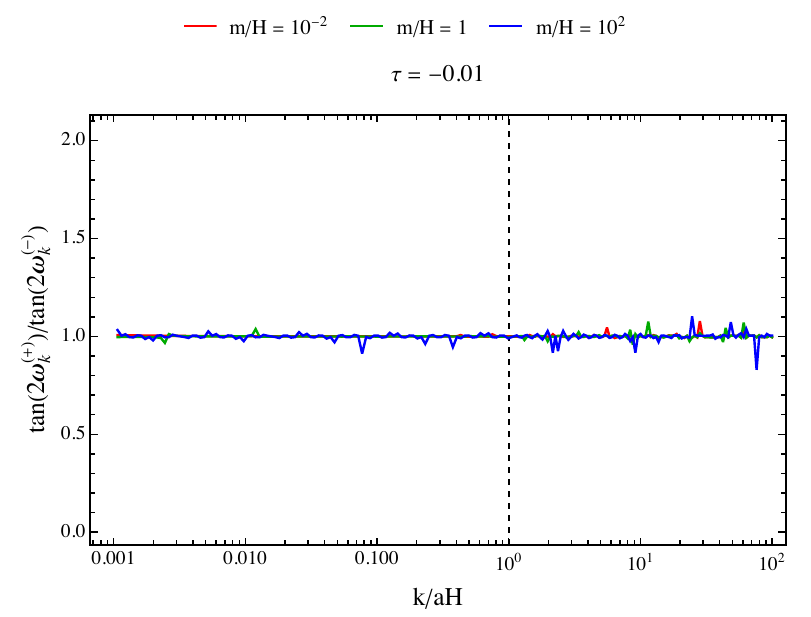}
 	\caption{Comparison of the squeezing parameters in the two helicity sectors in the small-$\vartheta$ regime.  The near coincidence of the $s=+\frac{1}{2}$ and $s=-\frac{1}{2}$ amplitudes implies that the partial-transpose spectra in \eqref{PartTransRhoSpSp} and \eqref{PartTransRhoSpSm} give almost identical logarithmic negativities. This justifies using the same-helicity expression \eqref{LogNegaSpSp} as the representative diagnostic in the numerical analysis.}
 		\label{ShowEqualBetweenPPwithPM}
 	\end{center}
 \end{figure}
\begin{figure}[ht]
 	\begin{center}
 	\includegraphics[scale=0.64]{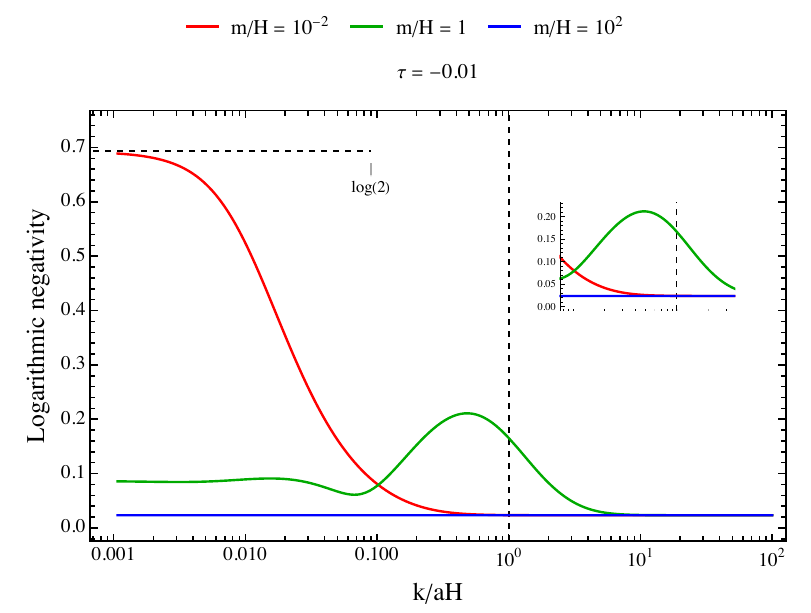}
 	\caption{Logarithmic negativity \eqref{LogNegaSpSp} for the same parameter choices as in Fig.~\ref{VonNeumanEntropyForPriMajorana}.  The curve tracks the inseparability of the Majorana pair and approaches the fermionic upper bound $\ln 2$ when the occupation number is driven close to $\mathcal N_k=1/2$.  The inset highlights the enhanced super-horizon signal for light Majorana modes.}
 		\label{LogNegatiForPriMajorana}
 	\end{center}
 \end{figure}

We next quantify the bipartite entanglement using the logarithmic negativity. This quantity is based on the positive-partial-transpose criterion: separable states remain positive under partial transposition, whereas a negative eigenvalue of the partially transposed density matrix signals non-separability \cite{Peres:1996dw,Horodecki:1996nc}. Vidal and Werner introduced the negativity as a computable mixed-state entanglement measure, and the logarithmic negativity is its additive form, which is widely used because it is straightforward to evaluate and bounds distillable entanglement \cite{Vidal:2002zz,Plenio:2005cwa}. In the present Majorana system it is especially useful because it directly measures the inseparability of the $\boldsymbol{k}$ and $-\boldsymbol{k}$ sectors, rather than only the mixedness of a reduced density matrix. For representative examples, we consider the helicity combinations $s=s^{\prime}=+\frac{1}{2}$ and $s=+\frac{1}{2},\,s^{\prime}=-\frac{1}{2}$. In the basis convention \eqref{TwoModeBasisConvention}, the corresponding density matrices can be written as
\begin{widetext}
\begin{align}
&\hat{\rho}_{\boldsymbol{k},-\boldsymbol{k}}^{(s=+\frac{1}{2};s^{\prime}=+\frac{1}{2})}=\left(\begin{array}{cccc}
c_{r_{k}^{+}}^{2} & 0 & 0 & -\text{i}\text{e}^{\text{i}\varphi_{k}^{(+)}}s_{r_{k}^{+}}c_{r_{k}^{+}}\\
0 & 0 & 0 & 0\\
0 & 0 & 0 & 0\\
\text{i}\text{e}^{-\text{i}\varphi_{k}^{(+)}}s_{r_{k}^{+}}c_{r_{k}^{+}} & 0 & 0 & s_{r_{k}^{+}}^{2}
\end{array}\right),
\end{align}and\begin{align}
&\hat{\rho}_{\boldsymbol{k},-\boldsymbol{k}}^{(s=+\frac{1}{2};s^{\prime}=-\frac{1}{2})}=\left(\begin{array}{cccc}
c_{r_{k}^{+}}c_{r_{k}^{-}}\text{e}^{\text{i}\Delta_{k}^{(1)}} & 0 & 0 & -\text{i}c_{r_{k}^{+}}s_{r_{k}^{-}}\text{e}^{\text{i}\Delta_{k}^{(2)}}\\
0 & 0 & 0 & 0\\
0 & 0 & 0 & 0\\
\text{i}s_{r_{k}^{+}}c_{r_{k}^{-}}\text{e}^{-\text{i}\bar{\Delta}_{k}^{(2)}} & 0 & 0 & s_{r_{k}^{+}}s_{r_{k}^{-}}\text{e}^{\text{i}\Delta_{k}^{(3)}}
\end{array}\right),
\end{align}
\end{widetext}
where we have denoted
\begin{align}
\nonumber
&\Delta_{k}^{(1)}=\omega_{k}^{(-)}-\omega_{k}^{(+)}~,~\Delta_{k}^{(2)}=\varphi_{k}^{(-)}+\omega_{k}^{(-)}-\omega_{k}^{(+)},\\
\nonumber
&\Delta_{k}^{(3)}=\varphi_{k}^{(-)}+\omega_{k}^{(-)}-\varphi_{k}^{(+)}-\omega_{k}^{(+)},\\
\nonumber
&\bar{\Delta}_{k}^{(2)}=\varphi_{k}^{(+)}+\omega_{k}^{(+)}-\omega_{k}^{(-)},
\end{align}and
\begin{align}
&c_{r_k^{+}}=\cos (r_k^{(+)}),c_{r_k^{-}}=\cos (r_k^{(-)}),\\
&s_{r_k^{+}}=\sin (r_k^{(+)}),s_{r_k^{-}}=\sin (r_k^{(-)}).
\end{align}For a density matrix
$\hat{\rho}_{\mathbf{k},-\mathbf{k}}$ associated with the two subsystems
$\mathbf{k}$ and $-\mathbf{k}$, we define
\begin{equation}
    E_{\mathcal N}\!\left[\hat{\rho}_{\mathbf{k},-\mathbf{k}}\right]
    \equiv
    \ln \left\|
        \hat{\rho}_{\mathbf{k},-\mathbf{k}}^{\,T_{-\mathbf{k}}}
    \right\|_1
    =
    \ln \left(1+2\sum_{\lambda_i<0}|\lambda_i|\right),
    \label{eq:logneg_definition}
\end{equation}
where $T_{-\mathbf{k}}$ denotes the partial transpose with respect to
the $-\mathbf{k}$ subsystem, $\|\cdots\|_1$ is the trace norm, and
$\{\lambda_i\}$ are the eigenvalues of
$\hat{\rho}_{\mathbf{k},-\mathbf{k}}^{\,T_{-\mathbf{k}}}$.  Applying this partial transpose to the two representative density matrices gives
 \begin{widetext}
\begin{align}
\label{PartTransRhoSpSp}
&\big(\hat{\rho}_{\boldsymbol{k},-\boldsymbol{k}}^{(s=+\frac{1}{2};s^{\prime}=+\frac{1}{2})}\big)^{T_{-\boldsymbol{k}}}=\left(\begin{array}{cccc}
(c_{r_k^{+}})^{2} & 0 & 0 & 0\\
0 & 0 & -\text{i}s_{r_k^{+}}c_{r_k^{+}}\text{e}^{\text{i}\varphi_{k}^{(+)}} & 0\\
0 & \text{i}s_{r_k^{+}}c_{r_k^{+}}\text{e}^{-\text{i}\varphi_{k}^{(+)}} & 0 & 0\\
0 & 0 & 0 & (c_{r_k^{+}})^{2}
\end{array}\right),
 \end{align}and 
\begin{align} 
\label{PartTransRhoSpSm} 
&\big( \hat{\rho}_{\boldsymbol{k},-\boldsymbol{k}}^{(s=+\frac{1}{2};s^{\prime}=-\frac{1}{2})}\big)^{T_{-\boldsymbol{k}}}\!=\!\!\!\left(\begin{array}{cccc}
c_{r_k^{+}}c_{r_k^{-}}\text{e}^{\text{i}\Delta_{k}^{(1)}} & 0 & 0 & 0\\
0 & 0 & -\text{i}c_{r_k^{+}}s_{r_k^{-}}\text{e}^{\text{i}\Delta_{k}^{(2)}} & 0\\
0 & \text{i}s_{r_k^{+}}c_{r_k^{-}}\text{e}^{-\text{i}\bar{\Delta}_{k}^{(2)}} & 0 & 0\\
0 & 0 & 0 & s_{r_k^{+}}s_{r_k^{-}}\text{e}^{\text{i}\Delta_{k}^{(3)}}
\end{array}\right), \end{align}
\end{widetext}
The eigenvalues corresponding to \eqref{PartTransRhoSpSp} and \eqref{PartTransRhoSpSm} are respectively
\begin{align}
\label{EigenPartTransRhoSpSp}
&\cos(r_{k}^{(+)})^{2}\,,\,\pm\cos(r_{k}^{(+)})\sin(r_{k}^{(+)})\,,\,\sin(r_{k}^{(+)})^{2},
\end{align}
and
\begin{align}
\nonumber
&\text{e}^{\text{i}\Delta_{k}^{(1)}}\cos(r_{k}^{(-)})\cos(r_{k}^{(+)}),\text{e}^{\text{i}\Delta_{k}^{(3)}}\sin(r_{k}^{(-)})\sin(r_{k}^{(+)}),\\
\label{EigenPartTransRhoSpSm}
&\mp\frac{1}{2}\text{e}^{\frac{\text{i}}{2}(\Delta_{k}^{(3)}+\Delta_{k}^{(1)})}\sqrt{\sin(2r_{k}^{(-)})\sin(2r_{k}^{(+)})}
\end{align}
Together with \eqref{SqueezeAmplitude}--\eqref{SqueezeRotationAngles}, Fig.~\ref{ShowEqualBetweenPPwithPM} shows that the two helicity sectors have almost identical squeezing amplitudes in the small-$\vartheta$ limit.  The phase factors in \eqref{EigenPartTransRhoSpSm} therefore do not change the physical conclusion, and the negative eigenvalues in \eqref{EigenPartTransRhoSpSp}--\eqref{EigenPartTransRhoSpSm} lead to
\begin{align}
\nonumber
E_{\mathcal{N}}^{(+)(+)}&=\ln\vert\vert\big(\hat{\rho}_{\boldsymbol{k},-\boldsymbol{k}}^{(s=+\frac{1}{2};s^{\prime}=+\frac{1}{2})}\big)^{T_{-\boldsymbol{k}}}\vert\vert_{1}\\
\label{LogNegaSpSp}
&=\ln\big(1+\sin(2r_{k}^{(+)})\big),
\end{align}and 
\begin{align}
\nonumber
E_{\mathcal{N}}^{(+)(-)}&=\ln\vert\vert\big(\hat{\rho}_{\boldsymbol{k},-\boldsymbol{k}}^{(s=+\frac{1}{2};s^{\prime}=-\frac{1}{2})}\big)^{T_{-\boldsymbol{k}}}\vert\vert_{1}\\
\nonumber
&=\ln\big(1\!+\!\sqrt{\sin(2r_{k}^{(-)})\sin(2r_{k}^{(+)})}\big)\\
\label{LogNegaSpSm}
&\approx\!E_{\mathcal{N}}^{(+)(+)}.
\end{align}
The logarithmic negativity therefore carries the same qualitative message as the von Neumann entropy, but it does so through the stronger and more operational criterion of bipartite inseparability.  The fermionic Majorana pair is bounded by $E_{\mathcal N}\leq \ln 2$, because the Grassmann nature of the field and the Pauli principle allow only the two occupation sectors $|0_{\mathbf{k}}0_{-\mathbf{k}}\rangle$ and $|1_{\mathbf{k}}1_{-\mathbf{k}}\rangle$.  Strong super-horizon quantumness is reached not through unbounded particle production, but by driving the occupation toward $\mathcal N_k\simeq1/2$, where the two sectors form an almost maximally inseparable pair.  This agrees with the entropy result in Fig.~\ref{VonNeumanEntropyForPriMajorana}: for small $m/H$, the Majorana modes retain enhanced super-horizon quantum correlations.  Similar uses of entanglement diagnostics in inflationary settings have emphasized the persistence and evolution of quantum correlations in cosmological perturbations \cite{Brahma:2020zpk,Brahma:2024ycc}; our result adds a fermionic, Pauli-bounded realization in which logarithmic negativity makes the inseparability of the produced pair especially explicit.  The scale dependence of this statement is displayed in Fig.~\ref{LogNegatiForPriMajorana}.  For the same parameter set used in Fig.~\ref{VonNeumanEntropyForPriMajorana}, the logarithmic negativity increases in the super-horizon regime for light Majorana modes and approaches the fermionic upper bound when the occupation number is driven close to $\mathcal N_k=1/2$.  Thus Fig.~\ref{LogNegatiForPriMajorana} confirms, through an inseparability measure rather than through reduced-state entropy, that the light-mass regime carries the strongest residual quantum correlation.

\begin{figure}[ht]
 	\begin{center}         
    \includegraphics[scale=0.55]{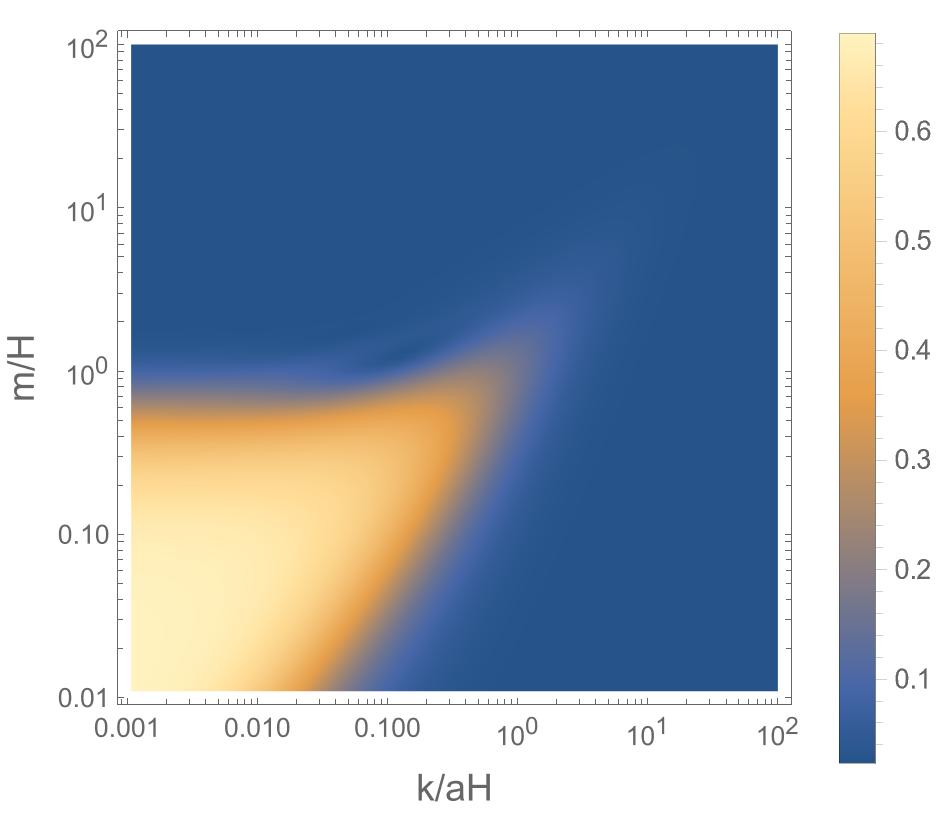}         \includegraphics[scale=0.95]{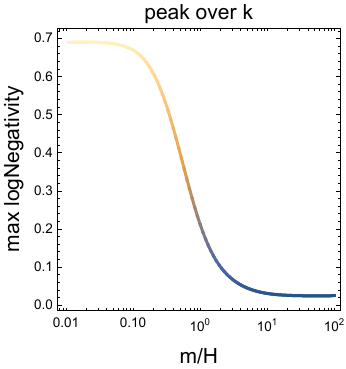}
 	\caption{Mass and scale dependence of the logarithmic negativity.  The upper panel shows $E_{\mathcal{N}}^{(+)(+)}$ in the $(m/H,k/aH)$ plane, while the lower panel extracts the maximal value over the super-horizon range for each fixed $m/H$.  The enhancement at small $m/H$ confirms that light Majorana modes carry stronger super-horizon bipartite quantumness.}
 		\label{HeatMapForMaxLogNegati}
 	\end{center}
 \end{figure}

Figure~\ref{HeatMapForMaxLogNegati} further summarizes the mass and scale dependence of the logarithmic negativity.  The heat map shows that the enhancement is concentrated in the small $m/H$ and super-horizon region, while the lower panel makes the suppression with increasing $m/H$ explicit by extracting the maximal super-horizon value.  This behavior supports the interpretation that light Majorana modes preserve a stronger two-mode quantum memory, whereas heavier modes are less efficient at maintaining near-maximal inseparability after horizon exit.

\section{Conclusion and discussion \label{ConcluDiscuss}}

In this work we used a primordial Majorana field as a fermionic probe of quantum correlations during inflation. Starting from the two-component spinor action on a torsion-free FLRW background, we derived the helicity mode equations with the Bunch--Davies initial condition and constructed the quadratic Hamiltonian in the paired momentum basis. We then showed that two descriptions of the time-dependent fermionic system, namely Hamiltonian diagonalization and the fermionic squeezing formalism, lead to the same Bogoliubov transformation of the creation and annihilation operators. This provides a compact technical route from the Majorana mode functions to the instantaneous occupation number and to the two-mode state for each pair $(\boldsymbol{k},-\boldsymbol{k})$.

The resulting state is simple but informative. Because of Fermi statistics, each helicity sector contains only the vacuum and one-pair sectors, so the relevant Hilbert space is finite and the occupation number is bounded. This allowed us to compute two complementary quantum-information diagnostics: the von Neumann entropy of the reduced mode and the logarithmic negativity of the bipartite state. Both measures indicate that light Majorana modes can retain enhanced super-horizon quantum correlations. In particular, the logarithmic negativity reaches its fermionic upper bound $E_{\mathcal N}=\ln 2$ when the occupation approaches $\mathcal{N}_{k}=1/2$, making explicit that the enhancement is not due to unbounded particle production, but to the inseparability of the two allowed occupation sectors. In this paper we have worked in the small-$\vartheta$ limit, which suppresses the effects of the cubic interaction $\partial_\mu \delta\phi\,\tilde{\chi}^{\dagger}\bar{\sigma}^{\mu}\tilde{\chi}$ generated by inflaton fluctuations around the homogeneous background. If $\vartheta$ is taken larger, this Yukawa-like interaction can itself provide an environmental coupling and its decohering effect should be included; a recent open-system analysis of scalar-fermion Yukawa theory in de Sitter shows explicitly how such interactions can generate entropy and decoherence at one loop \cite{Bhattacharya:2023twz}.

This result should be viewed in relation to several existing lines of work.  Studies of squeezed inflationary perturbations and environmental decoherence have shown how scalar and tensor modes can acquire an effectively classical description after horizon exit, but they also emphasize that classicality depends on the physical environment, coarse graining, and the observables being tested \cite{Polarski:1995jg,Kiefer:1998qe,Martin:2012pea,Burgess:2006jn,Martin:2018lin,Burgess:2022nwu,Sou:2022nsd,Hsiang:2021kgh}. The cosmological Bell program, including Maldacena's model with cosmological Bell inequalities, asks a closely related but more operational question: whether one can construct observables that certify the quantum origin of primordial correlations \cite{Campo:2005sv,Lim:2014uea,Maldacena:2015bha,Martin:2015qta,Martin:2017zxs,Sou:2024tjv}. Our analysis is more modest than a Bell-test proposal, but it is also more specific at the level of the state: it isolates a Pauli-bounded Majorana two-mode sector and tracks its entanglement directly. It also complements previous studies of fermionic entanglement and Bell-type diagnostics in de Sitter backgrounds \cite{Kanno:2016qcc,Bhattacharya:2020bal,Ali:2021jch} by focusing on primordial Majorana modes produced during axion inflation.

The main implication is therefore not that super-horizon matter fields must remain observable quantum objects. Rather, within the closed two-mode Majorana sector studied here, horizon exit alone does not automatically remove bipartite quantumness, especially for sufficiently light Majorana fields. If such correlations are to be classicalized by the end of inflation, the mechanism responsible for this loss of quantum coherence must be specified. This shifts the question from whether super-horizon stretching can make fields look classical in an effective sense to how a concrete decoherence channel, for example reheating interactions, inflaton-mediated couplings, gravitational nonlinearities, Yukawa interactions, or a thermalizing spectator sector, acts on the fermionic two-mode entanglement.

Several extensions are natural. First, the present closed-system calculation should be embedded in an open-system treatment, using an influence-functional or master-equation approach, to determine whether the logarithmic negativity survives realistic end-of-inflation and reheating dynamics \cite{Calzetta:2008iqa,Breuer:2002pc,Sieberer:2015svu,Prosen:2008bbr,Bhattacharya:2023twz}. Second, it would be useful to study primordial-black-hole (PBH) production scenarios in which a transient non-attractor phase is inserted between two attractor phases. Such backgrounds are designed to amplify curvature perturbations through a temporary departure from the slow-roll attractor, and the same enhanced time dependence may leave a sharper imprint on the Majorana squeezing parameter \cite{Kinney:2005vj,Motohashi:2017kbs,Inomata:2018cht,Di:2017ndc,Lin:2021vwc,Wang:2024euw}. Finally, one can ask whether the fermionic quantum memory identified here has any indirect observational footprint \cite{Adshead:2018oaa,Tong:2023krn}, or whether it is completely erased by the same interactions that populate the post-inflationary plasma \cite{Kofman:1997yn,Amin:2014eta}. Answering this question would turn the Majorana probe from a technical diagnostic of primordial quantumness into a controlled test of which decoherence mechanisms actually complete the quantum-to-classical transition.

\section{Acknowledgments}
A.-C. acknowledges support from the China Scholarship Council (CSC) under Grant No. 202008620074. A.-C. and H.-Q. were supported by the National Natural Science Foundation of China under Grant No. 12475104. K.W. acknowledges support from the CSC under Grant No. 202206540010, Grant No. PID2022-126224NB-C21, and the Generalitat de Catalunya under Grant No. 2021-SGR-249.

\appendix

\begin{widetext}

\section{Conventions for Gamma Matrices in Flat Spacetime and Useful Orthonormality Relations for Helicity Eigenspinors \label{SpinorInFlatSpacetime}}

We use the convention $\eta_{ab}=\text{diag}(+,-,-,-)$ in this appendix, following the spinor conventions of Refs.~\cite{Dreiner:2008tw,Lee:2026fpg}. The gamma matrices satisfy \begin{align}
\label{BasicGammaProperty}
&\{\gamma_{a},\gamma_{b}\}=2\eta_{ab}I_{4\times4}~,~(\gamma_{a})^{\dagger}=\boldsymbol{A}\gamma_{a}\boldsymbol{A}^{-1}\,,\,\boldsymbol{A}=\boldsymbol{A}^{-1}=\gamma_{0},
\end{align}Throughout this appendix we use the chiral representation, equivalently the Weyl basis,
\begin{align}
&\gamma^{a}=\left(\begin{array}{cc}
0 & \sigma^{a}\\
\bar{\sigma}^{a} & 0
\end{array}\right),\quad\gamma_{5}=\text{i}\gamma^{0}\gamma^{1}\gamma^{2}\gamma^{3}=\left(\begin{array}{cc}
-I_{2\times2} & 0\\
0 & I_{2\times2}
\end{array}\right).
\end{align}with the corresponding two-dimensional sigma matrices
\begin{align}
&\sigma^{a}=\{I_{2\times2},\boldsymbol{\sigma}\}~,~\bar{\sigma}^{a}=\{I_{2\times2},-\boldsymbol{\sigma}\}.
\end{align}In this representation, a four-component Dirac spinor $\Psi(x)$ is decomposed into two two-component spinors, $\chi(x)$ and $\xi(x)$, which carry opposite $U(1)$ charges. We choose the spinor-index convention
\begin{align}
\label{ChiralRepreFourComponentPsi}
&\left\lceil \Psi(x)\right\rceil _{(\boldsymbol{\frac{1}{2}},\boldsymbol{0})\oplus(\boldsymbol{0},\boldsymbol{\frac{1}{2}})}\equiv\left(\begin{array}{c}
\chi_{\mathrm{I}}(x)\\
\xi^{\dagger\dot{\mathrm{I}}}(x)
\end{array}\right),
\end{align}Here the undotted indices $\mathrm{I},\mathrm{J}=1,2$ label the left-handed $(\boldsymbol{\frac{1}{2}},\boldsymbol{0})$ representation, while the dotted indices $\dot{\mathrm{I}},\dot{\mathrm{J}}=\dot{1},\dot{2}$ label the right-handed $(\boldsymbol{0},\boldsymbol{\frac{1}{2}})$ representation. These two types of indices distinguish the two inequivalent irreducible components of a Dirac spinor under the Lorentz group. With the convention in \eqref{ChiralRepreFourComponentPsi}, the gamma matrices, $\gamma_5$, and the identity matrix have the following spinor-index structure:
\begin{align}
\label{GammaChiralRepSpinorIndice}
&\left\lceil \gamma^{a}\right\rceil _{(\boldsymbol{\frac{1}{2}},\boldsymbol{0})\oplus(\boldsymbol{0},\boldsymbol{\frac{1}{2}})}\!\equiv\!\left(\begin{array}{cc}
(\boldsymbol{0})_{\mathrm{I}}^{~\mathrm{J}} & (\sigma^{a})_{\mathrm{I}\dot{\mathrm{J}}}\\
(\bar{\sigma}^{a})^{\dot{\mathrm{I}}\mathrm{J}} & (\boldsymbol{0})_{~\dot{\mathrm{J}}}^{\dot{\mathrm{I}}}
\end{array}\right)\,,\,\left\lceil \gamma_{5}\right\rceil _{(\boldsymbol{\frac{1}{2}},\boldsymbol{0})\oplus(\boldsymbol{0},\boldsymbol{\frac{1}{2}})}\!\equiv\!\left(\begin{array}{cc}
-\delta_{\mathrm{I}}^{~\mathrm{J}} & (\boldsymbol{0})_{\mathrm{I}\dot{\mathrm{J}}}\\
(\boldsymbol{0})^{\dot{\mathrm{I}}\mathrm{J}} & \delta_{~\dot{\mathrm{J}}}^{\dot{\mathrm{I}}}
\end{array}\right),
\end{align}and
\begin{align}
&\left\lceil I_{4\times4}\right\rceil _{(\boldsymbol{\frac{1}{2}},\boldsymbol{0})\oplus(\boldsymbol{0},\boldsymbol{\frac{1}{2}})}\!\equiv\!\left(\begin{array}{cc}
\delta_{\mathrm{I}}^{~\mathrm{J}} & (\boldsymbol{0})_{\mathrm{I}\dot{\mathrm{J}}}\\
(\boldsymbol{0})^{\dot{\mathrm{I}}\mathrm{J}} & \delta_{~\dot{\mathrm{J}}}^{\dot{\mathrm{I}}}
\end{array}\right).
\end{align}The matrix $\boldsymbol{A}$ in \eqref{BasicGammaProperty} is equal to $\gamma^0$ in this representation, but it is useful to keep the notation $\boldsymbol{A}$ because it enters the definition of the Dirac conjugate:
\begin{align}
\label{SpinorIndiceDiracConjugate}
&\left\lceil \bar{\Psi}(x)\right\rceil _{(\boldsymbol{\frac{1}{2}},\boldsymbol{0})\oplus(\boldsymbol{0},\boldsymbol{\frac{1}{2}})}=\left\lceil \Psi^{\dagger}(x)\boldsymbol{A}\right\rceil _{(\boldsymbol{\frac{1}{2}},\boldsymbol{0})\oplus(\boldsymbol{0},\boldsymbol{\frac{1}{2}})}\equiv\left(\begin{array}{cc}
\xi^{\mathrm{I}}(x) & \chi_{\dot{\mathrm{I}}}^{\dagger}(x)\end{array}\right).
\end{align}This definition is fixed by the Lorentz-transformation law $\bar{\Psi}\to\bar{\Psi}\mathcal{M}^{-1}$, where $\mathcal{M}$ denotes the $4\times4$ spinor representation of the Lorentz group in $(\boldsymbol{\frac{1}{2}},\boldsymbol{0})\oplus(\boldsymbol{0},\boldsymbol{\frac{1}{2}})$. Equation~\eqref{SpinorIndiceDiracConjugate} then gives the spinor-index form of $\boldsymbol{A}$:
\begin{align}
&\left\lceil \boldsymbol{A}\right\rceil _{(\boldsymbol{\frac{1}{2}},\boldsymbol{0})\oplus(\boldsymbol{0},\boldsymbol{\frac{1}{2}})}\equiv\left(\begin{array}{cc}
(\boldsymbol{0})^{\dot{\mathrm{J}}\mathrm{I}} & \delta_{~~\dot{\mathrm{I}}}^{\dot{\mathrm{J}}}\\
\delta_{\mathrm{J}}^{~~\mathrm{I}} & (\boldsymbol{0})_{\mathrm{J}\dot{\mathrm{I}}}
\end{array}\right)\longleftrightarrow\left\lceil \boldsymbol{A}^{-1}\right\rceil _{(\boldsymbol{\frac{1}{2}},\boldsymbol{0})\oplus(\boldsymbol{0},\boldsymbol{\frac{1}{2}})}\equiv\left(\begin{array}{cc}
(\boldsymbol{0})_{\mathrm{J}\dot{\mathrm{I}}} & \delta_{\mathrm{J}}^{~~\mathrm{I}}\\
\delta_{~~\dot{\mathrm{I}}}^{\dot{\mathrm{J}}} & (\boldsymbol{0})^{\dot{\mathrm{J}}\mathrm{I}}
\end{array}\right).
\end{align}With these index conventions, the identity $(\gamma_{a})^{\dagger}=\boldsymbol{A}\gamma_{a}\boldsymbol{A}^{-1}$ can be checked directly at the level of spinor indices:
\begin{small}
\begin{align}
\nonumber
\big(\left\lceil \gamma^{a}\right\rceil _{(\boldsymbol{\frac{1}{2}},\boldsymbol{0})\oplus(\boldsymbol{0},\boldsymbol{\frac{1}{2}})}\big)^{\dagger}&\equiv\left(\begin{array}{cc}
(\boldsymbol{0})_{~\dot{\mathrm{I}}}^{\dot{\mathrm{J}}} & (\bar{\sigma}^{a})^{\dot{\mathrm{J}}\mathrm{I}}\\
(\sigma^{a})_{\mathrm{J}\dot{\mathrm{I}}} & (\boldsymbol{0})_{\mathrm{J}}^{~\mathrm{I}}
\end{array}\right)=\left(\begin{array}{cc}
(\boldsymbol{0})^{\dot{\mathrm{J}}\mathrm{I}_{1}} & \delta_{\,~\dot{\mathrm{I}}_{1}}^{\dot{\mathrm{J}}}\\
\delta_{\mathrm{J}}^{~\mathrm{I}_{1}} & (\boldsymbol{0})_{\mathrm{J}\dot{\mathrm{I}}_{1}}
\end{array}\right)\left(\begin{array}{cc}
(\boldsymbol{0})_{\mathrm{I}_{1}}^{~~\mathrm{J}_{2}} & (\sigma^{a})_{\mathrm{I}_{1}\dot{\mathrm{J}}_{2}}\\
(\bar{\sigma}^{a})^{\dot{\mathrm{I}}_{1}\mathrm{J}_{2}} & (\boldsymbol{0})_{~~\dot{\mathrm{J}}_{2}}^{\dot{\mathrm{I}}_{1}}
\end{array}\right)\left(\begin{array}{cc}
(\boldsymbol{0})_{\mathrm{J}_{2}\dot{\mathrm{I}}} & \delta_{\mathrm{J}_{2}}^{~~\mathrm{I}}\\
\delta_{~~\dot{\mathrm{I}}}^{\dot{\mathrm{J}}_{2}} & (\boldsymbol{0})^{\dot{\mathrm{J}}_{2}\mathrm{I}}
\end{array}\right)\\
\nonumber
&=\left\lceil \boldsymbol{A}\right\rceil _{(\boldsymbol{\frac{1}{2}},\boldsymbol{0})\oplus(\boldsymbol{0},\boldsymbol{\frac{1}{2}})}\cdot\left\lceil \gamma^{a}\right\rceil _{(\boldsymbol{\frac{1}{2}},\boldsymbol{0})\oplus(\boldsymbol{0},\boldsymbol{\frac{1}{2}})}\cdot\left\lceil \boldsymbol{A}^{-1}\right\rceil _{(\boldsymbol{\frac{1}{2}},\boldsymbol{0})\oplus(\boldsymbol{0},\boldsymbol{\frac{1}{2}})}.
\end{align}
\end{small}Using the notation introduced in \eqref{GammaChiralRepSpinorIndice}, the Lorentz generators in the spinor representation are written as
\begin{align}
&\left\lceil \gamma^{ab}\right\rceil _{(\boldsymbol{\frac{1}{2}},\boldsymbol{0})\oplus(\boldsymbol{0},\boldsymbol{\frac{1}{2}})}\!\equiv\!-2\text{i}\left(\begin{array}{cc}
(\sigma^{ab})_{\mathrm{I}}^{~\mathrm{J}} & (\boldsymbol{0})_{\mathrm{I}\dot{\mathrm{J}}}\\
(\boldsymbol{0})^{\dot{\mathrm{I}}\mathrm{J}} & (\bar{\sigma}^{ab})_{~\dot{\mathrm{J}}}^{\dot{\mathrm{I}}}
\end{array}\right),
\end{align}where $\sigma^{ab}$ and $\bar{\sigma}^{ab}$ are defined by
\begin{align}
&\begin{cases}
\!\begin{array}{c}
(\sigma^{ab})_{\mathrm{I}}^{~\mathrm{J}}=\frac{\text{i}}{4}\big((\sigma^{a})_{\mathrm{I}\dot{\mathrm{J}}_{1}}(\bar{\sigma}^{b})^{\dot{\mathrm{J}}_{1}\mathrm{J}}\!-\!(\sigma^{b})_{\mathrm{I}\dot{\mathrm{J}}_{1}}(\bar{\sigma}^{a})^{\dot{\mathrm{J}}_{1}\mathrm{J}}\big),\\
(\bar{\sigma}^{ab})_{~\dot{\mathrm{J}}}^{\dot{\mathrm{I}}}=\frac{\text{i}}{4}\big((\bar{\sigma}^{a})^{\dot{\mathrm{I}}\mathrm{J}_{1}}(\sigma^{b})_{\mathrm{J}_{1}\dot{\mathrm{J}}}-(\bar{\sigma}^{b})^{\dot{\mathrm{I}}\mathrm{J}_{1}}(\sigma^{a})_{\mathrm{J}_{1}\dot{\mathrm{J}}}\big).
\end{array}\end{cases}
\end{align}Note that, in Sec.~\ref{AxionInflationFermionModel}, we adopted the metric signature convention $\eta_{\tilde{a}\tilde{b}}=\mathrm{diag}(-,+,+,+)$ when constructing the Majorana-field Lagrangian in an FLRW background. Accordingly, to maintain consistency with the conventions for $\eta_{\tilde{a}\tilde{b}}$ and $\eta^{\tilde{a}\tilde{b}}$, we redefine
\begin{align}
\label{TransQFTsignsToGRsigns}
&\sigma^{\tilde{a}}=-\text{i}\sigma^{a}~,~\bar{\sigma}^{\tilde{a}}=-\text{i}\bar{\sigma}^{a}~,~\sigma_{\tilde{a}}=\text{i}\sigma_{a}~,~\bar{\sigma}_{\tilde{a}}=\text{i}\bar{\sigma}_{a}.
\end{align}In practical calculations it is convenient to use \eqref{TransQFTsignsToGRsigns} first, thereby removing the tilded local Lorentz indices before applying the $\sigma$-matrix identities listed in this appendix. The standard vierbein construction and the torsion-free spin connection are reviewed in Refs.~\cite{Freedman:2012zz,Lee:2025lyk,Li:2026cix}. For completeness, we record the spin-connection combinations that enter the FLRW Majorana action:
\begin{align}
\nonumber
&\omega_{\tau\tilde{a}\tilde{b}}(\sigma^{\tilde{a}\tilde{b}})_{\mathrm{I}}^{~\mathrm{J}}=0\,,\,\omega_{x\tilde{a}\tilde{b}}(\sigma^{\tilde{a}\tilde{b}})_{\mathrm{I}}^{~\mathrm{J}}=\frac{\text{i}a^{\prime}}{a}(\sigma^{0}\bar{\sigma}^{1})_{\mathrm{I}}^{~\mathrm{J}},\\
\nonumber
&\omega_{y\tilde{a}\tilde{b}}(\sigma^{\tilde{a}\tilde{b}})_{\mathrm{I}}^{~\mathrm{J}}=\frac{\text{i}a^{\prime}}{a}(\sigma^{0}\bar{\sigma}^{2})_{\mathrm{I}}^{~\mathrm{J}}\,,\,\omega_{z\tilde{a}\tilde{b}}(\sigma^{\tilde{a}\tilde{b}})_{\mathrm{I}}^{~\mathrm{J}}=\frac{\text{i}a^{\prime}}{a}(\sigma^{0}\bar{\sigma}^{3})_{\mathrm{I}}^{~\mathrm{J}},\\
&(e_{~\tilde{a}}^{\mu}\bar{\sigma}^{\tilde{a}})^{\dot{\mathrm{I}}_{1}\mathrm{J}_{1}}(\omega_{\mu\tilde{a}_{1}\tilde{b}_{1}}\sigma^{\tilde{a}_{1}\tilde{b}_{1}})_{\mathrm{J}_{1}}^{~~\mathrm{J}_{2}}=\frac{3a^{\prime}}{a^{2}}(\bar{\sigma}^{0})^{\dot{\mathrm{I}}_{1}\mathrm{J}_{2}}=\frac{3\text{i}a^{\prime}}{a^{2}}(\bar{\sigma}^{\tilde{0}})^{\dot{\mathrm{I}}_{1}\mathrm{J}_{2}}.\end{align}The only nonvanishing components of the torsion-free spin connection are
\begin{align}
&\omega_{x\tilde{1}\tilde{0}}\!=\!\omega_{y\tilde{2}\tilde{0}}\!=\!\omega_{z\tilde{3}\tilde{0}}\!=\!\frac{a^{\prime}}{a},~\omega_{x\tilde{0}\tilde{1}}\!=\!\omega_{y\tilde{0}\tilde{2}}\!=\!\omega_{z\tilde{0}\tilde{3}}\!=-\frac{a^{\prime}}{a}.
\end{align}
These relations play a central role in expanding the covariant Lagrangian \eqref{MajoranaAction} into a form suitable for explicit computations.

The mode-function decomposition \eqref{EigenSpinorDecomModev1}--\eqref{EigenSpinorDecomModev2} and the quadratic Hamiltonian constructed from it require two-component helicity eigenspinors, their completeness relations, and several spinor-index contraction identities. Before assigning explicit Lorentz-spinor indices, we regard these eigenspinors simply as a two-dimensional basis in momentum space. The two linearly independent helicity eigenspinors are
\begin{align}
\label{EigenSpinorVector}
&\xi_{(+\frac{1}{2})}=(\text{e}^{-\text{i}\phi/2}\cos\frac{\theta}{2},\text{e}^{\text{i}\phi/2}\sin\frac{\theta}{2})^{T}~,~\xi_{(-\frac{1}{2})}=(-\text{e}^{-\text{i}\phi/2}\sin\frac{\theta}{2},\text{e}^{\text{i}\phi/2}\cos\frac{\theta}{2})^{T},
\end{align}in which the angles $(\theta,\phi)$ corresponding to the directions of the unit momentum $\tilde{\boldsymbol{k}}_i=\frac{\boldsymbol{k}_i}{k}=\{\sin\theta\cos\phi,\sin\theta\sin\phi,\cos\theta\}$. Note that $\xi^\dagger_{(s)}(\tilde{\boldsymbol{k}})$ is obtained by Hermitian conjugation of \eqref{EigenSpinorVector}. At this stage, the helicity eigenspinors $\xi_{(s)}(\tilde{\boldsymbol{k}})$ and $\xi^\dagger_{(s)}(\tilde{\boldsymbol{k}})$ should be regarded merely as two-component basis spinors in helicity space. Their association with Lorentz-spinor indices is not intrinsic. Depending on the context, either $\xi_{(s)}(\tilde{\boldsymbol{k}})$ or $\xi^\dagger_{(s)}(\tilde{\boldsymbol{k}})$ may carry undotted indices $\mathrm{I},\mathrm{J}=1,2$ of the $(\boldsymbol{\frac{1}{2}},\boldsymbol{0})$ representation or dotted indices $\dot{\mathrm{I}},\dot{\mathrm{J}}=\dot{1},\dot{2}$ of the $(\boldsymbol{0},\boldsymbol{\frac{1}{2}})$ representation. Throughout this paper, the distinction will be made explicit whenever spinor indices are displayed. When acting the sigma matrices on the eigenspinors, the following relations will be useful
\begin{align}
&(\bar{\sigma}^{0})^{\dot{\mathrm{I}}_{1}\mathrm{J}_{1}}\xi_{s,\mathrm{J}_{1}}(\tilde{\boldsymbol{k}})=\xi_{s}^{\dot{\mathrm{I}}_{1}}(\tilde{\boldsymbol{k}})~,~(\bar{\sigma}^{0})^{\dot{\mathrm{I}}_{1}\mathrm{J}_{1}}\xi_{s,\mathrm{J}_{1}}^{\dagger}(\tilde{\boldsymbol{k}})=-\xi_{s}^{\dagger\dot{\mathrm{I}}_{1}}(\tilde{\boldsymbol{k}}),\\
&(\tilde{\boldsymbol{k}}_{i}\sigma^{i})_{\mathrm{I}_{1}\dot{\mathrm{J}}_{1}}\xi_{s}^{\dot{\mathrm{J}}_{1}}(\tilde{\boldsymbol{k}})=2s\xi_{s,\mathrm{I}_{1}}(\tilde{\boldsymbol{k}})~,~(\tilde{\boldsymbol{k}}_{i}\sigma^{i})_{\mathrm{I}_{1}\dot{\mathrm{J}}_{1}}\xi_{s}^{\dagger\dot{\mathrm{J}}_{1}}(\tilde{\boldsymbol{k}})=2s\xi_{s,\mathrm{I}_{1}}^{\dagger}(\tilde{\boldsymbol{k}}),\\
&(\tilde{\boldsymbol{k}}_{i}\bar{\sigma}^{i})^{\dot{\mathrm{I}}_{1}\mathrm{J}_{1}}\xi_{s,\mathrm{J}_{1}}(\tilde{\boldsymbol{k}})=-2s\xi_{s}^{\dot{\mathrm{I}}_{1}}(\tilde{\boldsymbol{k}})~,~(\tilde{\boldsymbol{k}}_{i}\bar{\sigma}^{i})^{\dot{\mathrm{I}}_{1}\mathrm{J}_{1}}\xi_{s,\mathrm{J}_{1}}^{\dagger}(\tilde{\boldsymbol{k}})=-2s\xi_{s}^{\dagger\dot{\mathrm{I}}_{1}}(\tilde{\boldsymbol{k}}),
\end{align}The corresponding helicity sums are
\begin{align}
\label{EigenSpinorHelicitySumv1}
&\sum_{s}\xi_{s,\mathrm{I}_{1}}(\tilde{\boldsymbol{k}})\xi_{s}^{\dagger\mathrm{J}_{1}}(\tilde{\boldsymbol{k}})=\delta_{\mathrm{I}_{1}}^{~~\mathrm{J}_{1}}~,~\sum_{s}\xi_{s}^{\dot{\mathrm{I}}_{1}}(\tilde{\boldsymbol{k}})\xi_{s,\dot{\mathrm{J}}_{1}}^{\dagger}(\tilde{\boldsymbol{k}})=\delta_{~~\dot{\mathrm{J}}_{1}}^{\dot{\mathrm{I}}_{1}},\\
&\sum_{s}\xi_{s,\mathrm{I}_{1}}(\tilde{\boldsymbol{k}})\xi_{s,\dot{\mathrm{J}}_{1}}^{\dagger}(\tilde{\boldsymbol{k}})=\sum_{s}\xi_{s,\mathrm{I}_{1}}(-\tilde{\boldsymbol{k}})\xi_{s,\dot{\mathrm{J}}_{1}}^{\dagger}(-\tilde{\boldsymbol{k}})=(\sigma_{0})_{\mathrm{I}_{1}\dot{\mathrm{J}}_{1}},\\
\label{EigenSpinorHelicitySumv2}
&\sum_{s}\xi_{s,\mathrm{I}_{1}}^{\dagger}(-\tilde{\boldsymbol{k}})\xi_{s,\dot{\mathrm{J}}_{1}}(-\tilde{\boldsymbol{k}})=\sum_{s}\xi_{s,\mathrm{I}_{1}}^{\dagger}(\tilde{\boldsymbol{k}})\xi_{s,\dot{\mathrm{J}}_{1}}(\tilde{\boldsymbol{k}})=(\sigma_{0})_{\mathrm{I}_{1}\dot{\mathrm{J}}_{1}},\\
\label{EigenSpinorHelicitySumv3}
&\xi_{+\frac{1}{2},\dot{\mathrm{I}}_{2}}^{\dagger}(\tilde{\boldsymbol{k}})(\bar{\sigma}^{0})^{\dot{\mathrm{I}}_{2}\mathrm{J}_{1}}\xi_{+\frac{1}{2},\mathrm{I}_{1}}(\tilde{\boldsymbol{k}})+\xi_{-\frac{1}{2},\dot{\mathrm{I}}_{2}}(-\tilde{\boldsymbol{k}})(\bar{\sigma}^{0})^{\dot{\mathrm{I}}_{2}\mathrm{J}_{1}}\xi_{-\frac{1}{2},\mathrm{I}_{1}}^{\dagger}(-\tilde{\boldsymbol{k}})=\delta_{\mathrm{I}_{1}}^{\mathrm{J}_{1}},\\
\label{EigenSpinorHelicitySumv4}
&\xi_{+\frac{1}{2},\dot{\mathrm{I}}_{2}}(-\tilde{\boldsymbol{k}})(\bar{\sigma}^{0})^{\dot{\mathrm{I}}_{2}\mathrm{J}_{1}}\xi_{+\frac{1}{2},\mathrm{I}_{1}}^{\dagger}(-\tilde{\boldsymbol{k}})+\xi_{-\frac{1}{2},\dot{\mathrm{I}}_{2}}^{\dagger}(\tilde{\boldsymbol{k}})(\bar{\sigma}^{0})^{\dot{\mathrm{I}}_{2}\mathrm{J}_{1}}\xi_{-\frac{1}{2},\mathrm{I}_{1}}(\tilde{\boldsymbol{k}})=\delta_{\mathrm{I}_{1}}^{\mathrm{J}_{1}},\\
&\sum_{s=\pm\frac{1}{2}}\!\big(\xi_{s,\mathrm{I}_{1}}(\tilde{\boldsymbol{k}})\xi_{s,\mathrm{J}_{1}}^{\dagger}(\tilde{\boldsymbol{k}})\!+\!\xi_{s,\mathrm{J}_{1}}(-\tilde{\boldsymbol{k}})\xi_{s,\mathrm{I}_{1}}^{\dagger}(-\tilde{\boldsymbol{k}})\big)\!=\!\sum_{s=\pm\frac{1}{2}}\big(\xi_{s,\dot{\mathrm{I}}_{1}}^{\dagger}(\tilde{\boldsymbol{k}})\xi_{s}^{\dot{\mathrm{I}}_{2}}(\tilde{\boldsymbol{k}})\!+\!\xi_{s,\dot{\mathrm{I}}_{1}}(-\tilde{\boldsymbol{k}})\xi_{s}^{\dagger\dot{\mathrm{I}}_{2}}(-\tilde{\boldsymbol{k}})\big)\!=\!0.
\end{align}We also need the following contractions over spinor indices:
\begin{align}
&\xi_{(s)}^{\mathrm{I}_{1}}\!(-\tilde{\boldsymbol{k}})\xi_{(s^{\prime}),\mathrm{I}_{1}}\!(\tilde{\boldsymbol{k}})=-\xi_{(s)}^{\mathrm{I}_{1}}\!(\tilde{\boldsymbol{k}})\xi_{(s^{\prime}),\mathrm{I}_{1}}\!(-\tilde{\boldsymbol{k}})=\delta_{ss^{\prime}},\\
&\xi_{(s),\dot{\mathrm{I}}_{1}}^{\dagger}(\tilde{\boldsymbol{k}})\xi_{(s^{\prime})}^{\dagger\dot{\mathrm{I}}_{1}}(-\tilde{\boldsymbol{k}})=-\xi_{(s),\dot{\mathrm{I}}_{1}}^{\dagger}(-\tilde{\boldsymbol{k}})\xi_{(s^{\prime})}^{\dagger\dot{\mathrm{I}}_{1}}(\tilde{\boldsymbol{k}})=\delta_{ss^{\prime}},\\
&\xi_{s,\dot{\mathrm{I}}_{1}}(-\tilde{\boldsymbol{k}})\xi_{s^{\prime}}^{\dot{\mathrm{I}}_{1}}(\tilde{\boldsymbol{k}})=-\xi_{s,\dot{\mathrm{I}}_{1}}(\tilde{\boldsymbol{k}})\xi_{s^{\prime}}^{\dot{\mathrm{I}}_{1}}(-\tilde{\boldsymbol{k}})=-\delta_{ss^{\prime}},\\
&\xi_{s}^{\dagger\mathrm{I}_{1}}(\tilde{\boldsymbol{k}})\xi_{s^{\prime},\mathrm{I}_{1}}^{\dagger}(-\tilde{\boldsymbol{k}})=-\xi_{s}^{\dagger\mathrm{I}_{1}}(-\tilde{\boldsymbol{k}})\xi_{s^{\prime},\mathrm{I}_{1}}^{\dagger}(\tilde{\boldsymbol{k}})=-\delta_{ss^{\prime}},\\
&\xi_{s,\dot{\mathrm{I}}_{1}}^{\dagger}(\tilde{\boldsymbol{k}})(\tilde{\boldsymbol{k}}_{i}\bar{\sigma}^{i})^{\dot{\mathrm{I}}_{1}\mathrm{J}_{1}}\xi_{s^{\prime},\mathrm{J}_{1}}(\tilde{\boldsymbol{k}})=-2s\delta_{ss^{\prime}}~,~\xi_{s,\dot{\mathrm{I}}_{1}}^{\dagger}(\tilde{\boldsymbol{k}})(\tilde{\boldsymbol{k}}_{i}\bar{\sigma}^{i})^{\dot{\mathrm{I}}_{1}\mathrm{J}_{1}}\xi_{s^{\prime},\mathrm{J}_{1}}^{\dagger}(-\tilde{\boldsymbol{k}})=2s\delta_{ss^{\prime}},\\
&\xi_{s,\dot{\mathrm{I}}_{1}}(-\tilde{\boldsymbol{k}})(\tilde{\boldsymbol{k}}_{i}\bar{\sigma}^{i})^{\dot{\mathrm{I}}_{1}\mathrm{J}_{1}}\xi_{s^{\prime},\mathrm{J}_{1}}(\tilde{\boldsymbol{k}})=2s\delta_{ss^{\prime}}~,~\xi_{s,\dot{\mathrm{I}}_{1}}(-\tilde{\boldsymbol{k}})(\tilde{\boldsymbol{k}}_{i}\bar{\sigma}^{i})^{\dot{\mathrm{I}}_{1}\mathrm{J}_{1}}\xi_{s^{\prime},\mathrm{J}_{1}}^{\dagger}(-\tilde{\boldsymbol{k}})=-2s\delta_{ss^{\prime}},\\
&\xi_{s}^{\mathrm{I}_{1}}(-\tilde{\boldsymbol{k}})(\tilde{\boldsymbol{k}}_{i}\sigma^{i})_{\mathrm{I}_{1}\dot{\mathrm{J}}_{1}}\xi_{s^{\prime}}^{\dagger\dot{\mathrm{J}}_{1}}(-\tilde{\boldsymbol{k}})=2s\delta_{ss^{\prime}}~,~\xi_{s}^{\dagger\mathrm{I}_{1}}(\tilde{\boldsymbol{k}})(\tilde{\boldsymbol{k}}_{i}\sigma^{i})_{\mathrm{I}_{1}\dot{\mathrm{J}}_{1}}\xi_{s^{\prime}}^{\dot{\mathrm{J}}_{1}}(\tilde{\boldsymbol{k}})=2s\delta_{ss^{\prime}},\\
&\xi_{s}^{\mathrm{I}_{1}}(-\tilde{\boldsymbol{k}})(\tilde{\boldsymbol{k}}_{i}\sigma^{i})_{\mathrm{I}_{1}\dot{\mathrm{J}}_{1}}\xi_{s^{\prime}}^{\dot{\mathrm{J}}_{1}}(\tilde{\boldsymbol{k}})=2s\delta_{ss^{\prime}}~,~\xi_{s}^{\dagger\mathrm{I}_{1}}(\tilde{\boldsymbol{k}})(\tilde{\boldsymbol{k}}_{i}\sigma^{i})_{\mathrm{I}_{1}\dot{\mathrm{J}}_{1}}\xi_{s^{\prime}}^{\dagger\dot{\mathrm{J}}_{1}}(-\tilde{\boldsymbol{k}})=2s\delta_{ss^{\prime}},
\end{align}Together with these, the contractions involving $\sigma^0$ and $\bar{\sigma}^0$ are
\begin{align}
&\xi_{(s),\dot{\mathrm{I}}_{1}}^{\dagger}(\tilde{\boldsymbol{k}})(\bar{\sigma}^{0})^{\dot{\mathrm{I}}_{1}\mathrm{J}_{1}}\xi_{s^{\prime},\mathrm{J}_{1}}(\tilde{\boldsymbol{k}})=\xi_{(s),\dot{\mathrm{I}}_{1}}^{\dagger}(-\tilde{\boldsymbol{k}})(\bar{\sigma}^{0})^{\dot{\mathrm{I}}_{1}\mathrm{J}_{1}}\xi_{s^{\prime},\mathrm{J}_{1}}(-\tilde{\boldsymbol{k}})=\delta_{ss^{\prime}},\\
&\xi_{(s)}^{\mathrm{I}_{1}}(-\tilde{\boldsymbol{k}})(\sigma_{0})_{\mathrm{I}_{1}\dot{\mathrm{J}}_{1}}\xi_{(s^{\prime})}^{\dagger\dot{\mathrm{J}}_{1}}(-\tilde{\boldsymbol{k}})=\xi_{(s)}^{\mathrm{I}_{1}}(\tilde{\boldsymbol{k}})(\sigma_{0})_{\mathrm{I}_{1}\dot{\mathrm{J}}_{1}}\xi_{(s^{\prime})}^{\dagger\dot{\mathrm{J}}_{1}}(\tilde{\boldsymbol{k}})=\delta_{ss^{\prime}},\\
&\xi_{s,\dot{\mathrm{I}}_{1}}^{\dagger}(\tilde{\boldsymbol{k}})(\bar{\sigma}^{0})^{\dot{\mathrm{I}}_{1}\mathrm{J}_{1}}\xi_{s^{\prime},\mathrm{J}_{1}}^{\dagger}(-\tilde{\boldsymbol{k}})=-\xi_{s,\dot{\mathrm{I}}_{1}}^{\dagger}(-\tilde{\boldsymbol{k}})(\bar{\sigma}^{0})^{\dot{\mathrm{I}}_{1}\mathrm{J}_{1}}\xi_{s^{\prime},\mathrm{J}_{1}}^{\dagger}(\tilde{\boldsymbol{k}})=-\delta_{ss^{\prime}},\\
&\xi_{s,\dot{\mathrm{I}}_{1}}(-\tilde{\boldsymbol{k}})(\bar{\sigma}^{0})^{\dot{\mathrm{I}}_{1}\mathrm{J}_{1}}\xi_{s^{\prime},\mathrm{J}_{1}}^{\dagger}(-\tilde{\boldsymbol{k}})=\xi_{s,\dot{\mathrm{I}}_{1}}(\tilde{\boldsymbol{k}})(\bar{\sigma}^{0})^{\dot{\mathrm{I}}_{1}\mathrm{J}_{1}}\xi_{s^{\prime},\mathrm{J}_{1}}^{\dagger}(\tilde{\boldsymbol{k}})=\delta_{ss^{\prime}},\\
&\xi_{s,\dot{\mathrm{I}}_{1}}(-\tilde{\boldsymbol{k}})(\bar{\sigma}^{0})^{\dot{\mathrm{I}}_{1}\mathrm{J}_{1}}\xi_{s^{\prime},\mathrm{J}_{1}}(\tilde{\boldsymbol{k}})=-\xi_{s,\dot{\mathrm{I}}_{1}}(\tilde{\boldsymbol{k}})(\bar{\sigma}^{0})^{\dot{\mathrm{I}}_{1}\mathrm{J}_{1}}\xi_{s^{\prime},\mathrm{J}_{1}}(-\tilde{\boldsymbol{k}})=-\delta_{ss^{\prime}},\\
&\xi_{s}^{\mathrm{I}_{1}}(-\tilde{\boldsymbol{k}})(\sigma^{0})_{\mathrm{I}_{1}\dot{\mathrm{J}}_{1}}\xi_{s^{\prime}}^{\dagger\dot{\mathrm{J}}_{1}}(-\tilde{\boldsymbol{k}})=\xi_{s}^{\mathrm{I}_{1}}(\tilde{\boldsymbol{k}})(\sigma^{0})_{\mathrm{I}_{1}\dot{\mathrm{J}}_{1}}\xi_{s^{\prime}}^{\dagger\dot{\mathrm{J}}_{1}}(\tilde{\boldsymbol{k}})=\delta_{ss^{\prime}},\\
&\xi_{s}^{\dagger\mathrm{I}_{1}}(\tilde{\boldsymbol{k}})(\sigma^{0})_{\mathrm{I}_{1}\dot{\mathrm{J}}_{1}}\xi_{s^{\prime}}^{\dot{\mathrm{J}}_{1}}(\tilde{\boldsymbol{k}})=\xi_{s}^{\dagger\mathrm{I}_{1}}(-\tilde{\boldsymbol{k}})(\sigma^{0})_{\mathrm{I}_{1}\dot{\mathrm{J}}_{1}}\xi_{s^{\prime}}^{\dot{\mathrm{J}}_{1}}(-\tilde{\boldsymbol{k}})=\delta_{ss^{\prime}},\\
&\xi_{s}^{\mathrm{I}_{1}}(-\tilde{\boldsymbol{k}})(\sigma^{0})_{\mathrm{I}_{1}\dot{\mathrm{J}}_{1}}\xi_{s^{\prime}}^{\dot{\mathrm{J}}_{1}}(\tilde{\boldsymbol{k}})=-\xi_{s}^{\mathrm{I}_{1}}(\tilde{\boldsymbol{k}})(\sigma^{0})_{\mathrm{I}_{1}\dot{\mathrm{J}}_{1}}\xi_{s^{\prime}}^{\dot{\mathrm{J}}_{1}}(-\tilde{\boldsymbol{k}})=\delta_{ss^{\prime}},\\
&\xi_{s}^{\dagger\mathrm{I}_{1}}(\tilde{\boldsymbol{k}})(\sigma^{0})_{\mathrm{I}_{1}\dot{\mathrm{J}}_{1}}\xi_{s^{\prime}}^{\dagger\dot{\mathrm{J}}_{1}}(-\tilde{\boldsymbol{k}})=-\xi_{s}^{\dagger\mathrm{I}_{1}}(-\tilde{\boldsymbol{k}})(\sigma^{0})_{\mathrm{I}_{1}\dot{\mathrm{J}}_{1}}\xi_{s^{\prime}}^{\dagger\dot{\mathrm{J}}_{1}}(\tilde{\boldsymbol{k}})=\delta_{ss^{\prime}}.
\end{align}Finally, we introduce the antisymmetric tensors $\varepsilon^{\mathrm{I}\mathrm{J}}$, $\varepsilon^{\dot{\mathrm{I}}\dot{\mathrm{J}}}$, $\varepsilon_{\mathrm{I}\mathrm{J}}$, and $\varepsilon_{\dot{\mathrm{I}}\dot{\mathrm{J}}}$ as the spinor metrics:
\begin{align}
&\varepsilon_{\mathrm{I}\mathrm{J}}=\left(\begin{array}{cc}
0 & -1\\
+1 & 0
\end{array}\right)\sim-\text{i}\sigma^{2}~,~\varepsilon^{\dot{\mathrm{I}}\dot{\mathrm{J}}}=\left(\begin{array}{cc}
0 & +1\\
-1 & 0
\end{array}\right)\sim\text{i}\sigma^{2},\\
&\varepsilon^{\mathrm{I}\mathrm{J}}=\left(\begin{array}{cc}
0 & +1\\
-1 & 0
\end{array}\right)\sim\text{i}\sigma^{2}~,~\varepsilon_{\dot{\mathrm{I}}\dot{\mathrm{J}}}=\left(\begin{array}{cc}
0 & -1\\
+1 & 0
\end{array}\right)\sim-\text{i}\sigma^{2}.
\end{align}When these tensors are used to raise or lower spinor indices, we adopt the conventions \begin{align} &\varepsilon^{\mathrm{I}\mathrm{J}_{1}}\varepsilon_{\mathrm{J}_{1}\mathrm{J}}=-\varepsilon^{\mathrm{J}_{1}\mathrm{I}}\varepsilon_{\mathrm{J}_{1}\mathrm{J}}=\delta_{\mathrm{J}}^{\mathrm{I}}\,,\,\varepsilon_{\dot{\mathrm{I}}\dot{\mathrm{J}}_{1}}\varepsilon^{\dot{\mathrm{J}}_{1}\dot{\mathrm{J}}}=-\varepsilon_{\dot{\mathrm{J}}_{1}\dot{\mathrm{I}}}\varepsilon^{\dot{\mathrm{J}}_{1}\dot{\mathrm{J}}}=\delta_{\dot{\mathrm{I}}}^{\dot{\mathrm{J}}},\\ &\chi_{\mathrm{I}}=\varepsilon_{\mathrm{I}\mathrm{J}_{1}}\chi^{\mathrm{J}_{1}}\,,\,\chi^{\mathrm{I}}=\varepsilon_{\mathrm{I}\mathrm{J}_{1}}\chi^{\mathrm{J}_{1}}\,,\,\chi_{\dot{\mathrm{I}}}^{\dagger}=\varepsilon_{\dot{\mathrm{I}}\dot{\mathrm{J}}_{1}}\chi^{\dot{\mathrm{J}}_{1}}\,,\,\chi^{\dagger\dot{\mathrm{I}}}=\varepsilon^{\dot{\mathrm{I}}\dot{\mathrm{J}}_{1}}\chi_{\dot{\mathrm{J}}_{1}},\\ &\mathcal{A}^{\mathrm{I}\mathrm{J}}\!=\!\varepsilon^{\mathrm{I}\mathrm{I}_{1}}\varepsilon^{\mathrm{J}\mathrm{J}_{1}}\mathcal{A}_{\mathrm{I}_{1}\mathrm{J}_{1}}\,,\,\mathcal{A}_{\mathrm{I}\mathrm{J}}\!=\!\varepsilon_{\mathrm{I}\mathrm{I}_{1}}\varepsilon_{\mathrm{J}\mathrm{J}_{1}}\mathcal{A}^{\mathrm{I}_{1}\mathrm{J}_{1}}\,,\,\mathcal{A}^{\dot{\mathrm{I}}\dot{\mathrm{J}}}\!=\!\varepsilon^{\dot{\mathrm{I}}\dot{\mathrm{I}}_{1}}\varepsilon^{\mathrm{\dot{J}}\dot{\mathrm{J}}_{1}}\mathcal{A}_{\dot{\mathrm{I}}_{1}\dot{\mathrm{J}}_{1}}\,,\,\mathcal{A}_{\dot{\mathrm{I}}\dot{\mathrm{J}}}\!=\!\varepsilon_{\dot{\mathrm{I}}\dot{\mathrm{I}}_{1}}\varepsilon_{\dot{\mathrm{J}}\dot{\mathrm{J}}_{1}}\mathcal{A}^{\dot{\mathrm{I}}_{1}\dot{\mathrm{J}}_{1}}.\end{align}Note that the contraction patterns of the form $A^{\mathrm{I}\dots } B_{\mathrm{I}\dots}$ and $A_{\dot{\mathrm{I}}\dots } B^{\dot{\mathrm{I}}\dots}$ are generally spoiled once the spinor metric tensors are employed to raise or lower the spinor indices. Nevertheless, in all other circumstances we strictly adhere to the convention: namely, contractions for undotted indices are performed in the descending order, while for dotted indices they are performed in the ascending order.

\section{Derivation of the Mode-Function Equations of Motion and Their Solutions with the Bunch--Davies Vacuum as the Initial Condition \label{AppendixB}}

By varying the action \eqref{MajoranaActionConforTime} with respect to $\chi^\dagger_{\dot{\mathrm{I}}1}$, the equations of motion for the Majorana spinor field $\chi_{\mathrm{J}_{1}}(\tau,\boldsymbol{x})$ are obtained as
\begin{align}
\label{HalfCovaDiracEOMs}
&-(e_{~\tilde{a}}^{\mu}\bar{\sigma}^{\tilde{a}})^{\dot{\mathrm{I}}_{1}\mathrm{J}_{1}}\partial_{\mu}\chi_{\mathrm{J}_{1}}+\frac{\text{i}C_{\text{A-F}}}{f}\partial_{\mu}\bar{\phi}(e_{~\tilde{a}}^{\mu}\bar{\sigma}^{\tilde{a}})^{\dot{\mathrm{I}}_{1}\mathrm{J}_{1}}\chi_{\mathrm{J}_{1}}=m\chi^{\dagger,\dot{\mathrm{I}}_{1}}.
\end{align}We assume that a well-defined adiabatic vacuum can be specified at the initial time $\tau_0$. The field operator and its Hermitian conjugate are then expanded in terms of the creation and annihilation operators defined at $\tau_0$ as
\begin{align}
\label{ModeChi}
&\hat{\chi}_{\mathrm{I}_{1}}(\tau,\boldsymbol{x})=\sum_{s=\pm\frac{1}{2}}\int\frac{d^{3}\boldsymbol{k}}{(2\pi)^{3}}\big(x_{\mathrm{I}_{1},\boldsymbol{k}}^{(s)}(\tau)\text{e}^{\text{i}\boldsymbol{k}\cdot\boldsymbol{x}}\hat{a}_{\boldsymbol{k}}^{(s)}(\tau_{0})+y_{\mathrm{I}_{1},\boldsymbol{k}}^{(s)}(\tau)\text{e}^{-\text{i}\boldsymbol{k}\cdot\boldsymbol{x}}\hat{a}_{\boldsymbol{k}}^{(s)\dagger}(\tau_{0})\big),\\
\label{ModeChiDagger}
&\hat{\chi}_{\dot{\mathrm{I}}_{1}}^{\dagger}(\tau,\boldsymbol{x})=\sum_{s=\pm\frac{1}{2}}\int\frac{d^{3}\boldsymbol{k}}{(2\pi)^{3}}\big(x_{\dot{\mathrm{I}}_{1},\boldsymbol{k}}^{(s)\dagger}(\tau)\text{e}^{-\text{i}\boldsymbol{k}\cdot\boldsymbol{x}}\hat{a}_{\boldsymbol{k}}^{(s)\dagger}(\tau_{0})+y_{\dot{\mathrm{I}}_{1},\boldsymbol{k}}^{(s)\dagger}(\tau)\text{e}^{\text{i}\boldsymbol{k}\cdot\boldsymbol{x}}\hat{a}_{\boldsymbol{k}}^{(s)}(\tau_{0})\big).
\end{align}The spinor-valued mode functions are separated into scalar mode functions and helicity eigenspinors according to
\begin{align}
\label{EigenSpinorDecomModev1}
&x_{\mathrm{I}_{1},\boldsymbol{k}}^{(s)}(\tau)=X_{k}^{(s)}(\tau)\xi_{s,\mathrm{I}_{1}}(\tilde{\boldsymbol{k}})~,~y_{\boldsymbol{k}}^{\dagger(s),\dot{\mathrm{I}}_{1}}(\tau)=Y_{k}^{\star(s)}(\tau)\xi_{s}^{\dot{\mathrm{I}}_{1}}(\tilde{\boldsymbol{k}}),\\
\label{EigenSpinorDecomModev2}
&x_{\dot{\mathrm{I}}_{1},\boldsymbol{k}}^{(s)\dagger}(\tau)=X_{k}^{(s)\star}(\tau)\xi_{s,\dot{\mathrm{I}}_{1}}^{\dagger}(\tilde{\boldsymbol{k}})~,~y_{\boldsymbol{k}}^{(s),\mathrm{I}_{1}}(\tau)=Y_{k}^{(s)}(\tau)\xi_{s}^{\dagger\mathrm{I}_{1}}(\tilde{\boldsymbol{k}}).
\end{align}Substituting \eqref{EigenSpinorDecomModev1} and \eqref{EigenSpinorDecomModev2} into the covariant equation \eqref{HalfCovaDiracEOMs}, and using the helicity identities summarized in Appendix~\ref{SpinorInFlatSpacetime}, gives
\begin{small}
\begin{align}
&\frac{\text{i}}{a}X_{k}^{\prime\,(s)}(\tau)\underbrace{(\bar{\sigma}^{0})^{\dot{\mathrm{I}}_{1}\mathrm{J}_{1}}\xi_{s,\mathrm{J}_{1}}(\tilde{\boldsymbol{k}})}_{\xi_{s}^{\dot{\mathrm{I}}_{1}}(\tilde{\boldsymbol{k}})}-X_{k}^{(s)}(\tau)\frac{k}{a}\underbrace{(\tilde{\boldsymbol{k}}_{i}\bar{\sigma}^{i})^{\dot{\mathrm{I}}_{1}\mathrm{J}_{1}}\xi_{s,\mathrm{J}_{1}}(\tilde{\boldsymbol{k}})}_{-2s\xi_{s}^{\dot{\mathrm{I}}_{1}}(\tilde{\boldsymbol{k}})}+\frac{C_{\text{A-F}}\bar{\phi}^{\prime}}{af}X_{k}^{(s)}(\tau)\underbrace{(\bar{\sigma}^{0})^{\dot{\mathrm{I}}_{1}\mathrm{J}_{1}}\xi_{s,\mathrm{J}_{1}}(\tilde{\boldsymbol{k}})}_{\xi_{s}^{\dot{\mathrm{I}}_{1}}(\tilde{\boldsymbol{k}})}=mY_{k}^{\star(s)}(\tau)\xi_{s}^{\dot{\mathrm{I}}_{1}}(\tilde{\boldsymbol{k}}),\\
&\frac{\text{i}}{a}Y_{k}^{\prime\,(s)}(\tau)\underbrace{(\bar{\sigma}^{0})^{\dot{\mathrm{I}}_{1}\mathrm{J}_{1}}\xi_{s,\mathrm{J}_{1}}^{\dagger}(\tilde{\boldsymbol{k}})}_{-\xi_{s}^{\dagger\dot{\mathrm{I}}_{1}}(\tilde{\boldsymbol{k}})}+Y_{k}^{(s)}(\tau)\frac{k}{a}\underbrace{(\tilde{\boldsymbol{k}}_{i}\bar{\sigma}^{i})^{\dot{\mathrm{I}}_{1}\mathrm{J}_{1}}\xi_{s,\mathrm{J}_{1}}^{\dagger}(\tilde{\boldsymbol{k}})}_{-2s\xi_{s}^{\dagger\dot{\mathrm{I}}_{1}}(\tilde{\boldsymbol{k}})}+\frac{C_{\text{A-F}}\bar{\phi}^{\prime}}{af}Y_{k}^{(s)}(\tau)\underbrace{(\bar{\sigma}^{0})^{\dot{\mathrm{I}}_{1}\mathrm{J}_{1}}\xi_{s,\mathrm{J}_{1}}^{\dagger}(\tilde{\boldsymbol{k}})}_{-\xi_{s}^{\dagger\dot{\mathrm{I}}_{1}}(\tilde{\boldsymbol{k}})}=m\,X_{k}^{(s)\star}(\tau)\xi_{s}^{\dagger\dot{\mathrm{I}}_{1}}(\tilde{\boldsymbol{k}}).
\end{align}
\end{small}By factoring out the common helicity eigen-spinors, one can extract the following two sets of first-order differential equations, which are independent of the explicit eigen-spinor basis:
\begin{align}
\label{ModeFirstODEsForX}
&\text{i}X_{k}^{\prime\,(s)}(\tau)+\big(2sk+\frac{C_{\text{A-F}}\bar{\phi}^{\prime}(\tau)}{f}\big)X_{k}^{\,(s)}(\tau)=ma(\tau)Y_{k}^{\star(s)}(\tau),\\
\label{ModeFirstODEsForY}
&\text{i}Y_{k}^{\prime\,(s)}(\tau)+\big(2sk+\frac{C_{\text{A-F}}\bar{\phi}^{\prime}(\tau)}{f}\big)Y_{k}^{(s)}(\tau)=-ma(\tau)\,X_{k}^{(s)\star}(\tau).
\end{align}These equations can be decoupled into a common second-order equation. Denoting either scalar mode function by $Z_{k}^{(s)}(\tau)$, with $Z=X$ or $Z=Y$, one finds
\begin{align}
&Z_{k}^{\prime\prime\,(s)}(\tau)-\frac{a^{\prime}}{a}Z_{k}^{\prime\,(s)}(\tau)+\big\{ m^{2}a^{2}+\text{i}\frac{a^{\prime}}{a}(2sk+\frac{C_{\text{A-F}}}{f}\bar{\phi}^{\prime})+(2sk+\frac{C_{\text{A-F}}}{f}\bar{\phi}^{\prime})^{2}-\text{i}\frac{C_{\text{A-F}}}{f}\bar{\phi}^{\prime\prime}\big\} Z_{k}^{\,(s)}(\tau)=0.
\end{align}During slow-roll inflation, the homogeneous axion rolls at an approximately constant rate in cosmic time, $\ddot{\bar{\phi}}\ll H\dot{\bar{\phi}}$. This allows us to introduce the nearly constant parameter
\begin{align}
\label{AxionCouplingWithSRone}
&\text{const}=\vartheta=-\frac{C_{\text{A-F}}}{f}\frac{\dot{\bar{\phi}}}{H}=\frac{\sqrt{2\epsilon}C_{\text{A-F}}M_{\text{pl}}}{f},
\end{align}where $\epsilon$ is the first slow-roll parameter. Working in the quasi-de Sitter limit $a(\tau)=-\frac{1}{H\tau}$, and redefining $Z_{k}^{\,(s)}(\tau)=\sqrt{a(\tau)}\tilde{Z}_{k}^{\,(s)}(\tau)$ with $u=2\text{i}k\tau$, the mode equation becomes
\begin{align}
\label{DecoupleSecondOrderEOMs}
&\frac{d^{2}\tilde{Z}_{k}^{\,(s)}}{du^{2}}+\big(-\frac{1}{4}-\frac{2s(\text{i}\vartheta+\frac{1}{2})}{u}+\frac{1}{u^{2}}(\frac{1}{4}+\frac{m^{2}}{H^{2}}+\vartheta^{2})\big)\tilde{Z}_{k}^{\,(s)}=0.
\end{align}If we define $\lambda=-2s(\text{i}\vartheta+\frac{1}{2})$ and $\nu^{2}=-(\frac{m^{2}}{H^{2}}+\vartheta^{2})$, the general solution of \eqref{DecoupleSecondOrderEOMs} can be written in terms of Whittaker functions \cite{Adshead:2015kza} as
\begin{align}
&\tilde{Z}_{k}^{\,(s)}(\tau)=C_{1}^{(s)}(\text{i}k)\,W_{\lambda,\nu}(u)+C_{2}^{(s)}(\text{i}k)\,W_{-\lambda,\nu}(-u).
\end{align}The coefficients are fixed by the Bunch--Davies initial condition in the far past \cite{Chernikov:1968zm,Bunch:1978yq,Candelas:1975du,Kanno:2016qcc}. For $u=2\text{i}k\tau\to-\text{i}\infty$, the two Whittaker branches behave as $W_{\lambda,\nu}(u)\propto \text{e}^{-\text{i}k\tau}$ and $W_{-\lambda,\nu}(-u)\propto \text{e}^{+\text{i}k\tau}$, up to phase and power-law factors. The particle mode $X_{k}^{(s)}$ is therefore selected by setting $C_{2}^{(s)}(\text{i}k)=0$. For the antiparticle mode $Y_{k}^{(s)}$, the coupled first-order system \eqref{ModeFirstODEsForX}--\eqref{ModeFirstODEsForY} fixes the conjugate frequency assignment, so the Bunch--Davies choice sets $C_{1}^{(s)}(\text{i}k)=0$. Equivalently, $X_{k}^{(s)}$ must be paired consistently with $Y_{k}^{\star(s)}$, and $Y_{k}^{(s)}$ with $X_{k}^{\star(s)}$. More details on this Whittaker-function basis and the corresponding Bunch--Davies branch choice can be found in \cite{Adshead:2015kza,Cruces:2026qvl}. With these choices, the solutions can be written as \cite{Adshead:2015kza}
\begin{align}
\label{SolXkPlus}
X_{k}^{\,(+\frac{1}{2})}(\tau)&=-\frac{\text{i}m}{H}\frac{\text{e}^{\text{i}\theta}\text{e}^{-\frac{\pi}{2}\vartheta}}{\sqrt{2k\tau}}W_{-\text{i}\vartheta-\frac{1}{2},\text{i}\sqrt{\frac{m^{2}}{H^{2}}+\vartheta^{2}}}(2\text{i}k\tau),\\
\label{SolXkMinus}
X_{k}^{\,(-\frac{1}{2})}(\tau)&=\frac{\text{e}^{\text{i}\theta}\text{e}^{\frac{\pi}{2}\vartheta}}{\sqrt{2k\tau}}W_{\text{i}\vartheta+\frac{1}{2},\text{i}\sqrt{\frac{m^{2}}{H^{2}}+\vartheta^{2}}}(2\text{i}k\tau),\\
\label{SolYkPlus}
Y_{k}^{\star(+\frac{1}{2})}(\tau)&=\frac{\text{e}^{\text{i}\theta^{\prime}}\text{e}^{-\frac{\pi}{2}\vartheta}}{\sqrt{2k\tau}}W_{-\text{i}\vartheta+\frac{1}{2},\text{i}\sqrt{\frac{m^{2}}{H^{2}}+\vartheta^{2}}}(2\text{i}k\tau),\\
\label{SolYkMinus}
Y_{k}^{\star(-\frac{1}{2})}(\tau)&=-\frac{\text{i}m}{H}\frac{\text{e}^{\text{i}\theta^{\prime}}\text{e}^{\frac{\pi}{2}\vartheta}}{\sqrt{2k\tau}}W_{\text{i}\vartheta-\frac{1}{2},\text{i}\sqrt{\frac{m^{2}}{H^{2}}+\vartheta^{2}}}(2\text{i}k\tau).
\end{align}Here $\theta$ and $\theta^\prime$ are constant phase angles. They do not affect the physical observables considered in this work, such as the particle number and the parameters of the two-mode quantum state. The remaining normalization of the mode functions is fixed by canonical quantization. To derive the corresponding Wronskian condition explicitly, we first write the action \eqref{MajoranaActionConforTime} in the form
\begin{align}
&S_{\text{(Majo)}}\!=\!\int d^{4}x\big\{\text{i}\chi_{\dot{\mathrm{I}}_{1}}^{\dagger}(\bar{\sigma}^{0})^{\dot{\mathrm{I}}_{1}\mathrm{J}_{1}}\chi_{\mathrm{J}_{1}}^{\prime}+\text{i}\chi_{\dot{\mathrm{I}}_{1}}^{\dagger}(\bar{\sigma}^{i})^{\dot{\mathrm{I}}_{1}\mathrm{J}_{1}}\partial_{i}\chi_{\mathrm{J}_{1}}-\frac{a(\tau)m}{2}(\chi^{\mathrm{I}_{1}}\chi_{\mathrm{I}_{1}}+\chi_{\dot{\mathrm{I}}_{1}}^{\dagger}\chi^{\dagger,\dot{\mathrm{I}}_{1}})+\frac{C_{\text{A-F}}}{f}\bar{\phi}^{\prime}\chi_{\dot{\mathrm{I}}_{1}}^{\dagger}(\bar{\sigma}^{0})^{\dot{\mathrm{I}}_{1}\mathrm{J}_{1}}\chi_{\mathrm{J}_{1}}\big\}.
\end{align}The conjugate momentum is therefore
\begin{align}
&\pi_{\chi}^{\mathrm{J}_{1}}=\frac{\partial\mathcal{L}_{\text{(Majo)}}}{\partial\chi_{\mathrm{J}_{1}}^{\prime}}=\text{i}\chi_{\dot{\mathrm{I}}_{1}}^{\dagger}(\bar{\sigma}^{0})^{\dot{\mathrm{I}}_{1}\mathrm{J}_{1}},
\end{align}and its operator expansion is
\begin{align}
\label{ModeConjugatePi}
&\hat{\pi}_{\chi}^{\mathrm{J}_{1}}(\tau,\boldsymbol{x})=\text{i}\sum_{s=\pm\frac{1}{2}}\int\frac{d^{3}\boldsymbol{k}}{(2\pi)^{3}}\big(x_{\dot{\mathrm{I}}_{1},\boldsymbol{k}}^{(s)\dagger}(\tau)\text{e}^{-\text{i}\boldsymbol{k}\cdot\boldsymbol{x}}\hat{a}_{\boldsymbol{k}}^{(s)\dagger}(\tau_{0})+y_{\dot{\mathrm{I}}_{1},\boldsymbol{k}}^{(s)\dagger}(\tau)\text{e}^{\text{i}\boldsymbol{k}\cdot\boldsymbol{x}}\hat{a}_{\boldsymbol{k}}^{(s)}(\tau_{0})\big)(\bar{\sigma}^{0})^{\dot{\mathrm{I}}_{1}\mathrm{J}_{1}}.
\end{align}Canonical quantization imposes the equal-time anti-commutation relations
\begin{align}
\label{AntiCommutationDefine}
&\{\hat{\chi}_{\mathrm{I}_{1}}(\tau,\boldsymbol{x}),\hat{\pi}_{\chi}^{\mathrm{J}_{1}}(\tau,\boldsymbol{y})\}=\text{i}\delta_{\mathrm{I}_{1}}^{\mathrm{J}_{1}}\delta^{3}(\boldsymbol{x}-\boldsymbol{y})~,~\{\hat{a}_{\boldsymbol{k}}^{(s)}(\tau_{0}),\hat{a}_{\boldsymbol{q}}^{(s^{\prime})\dagger}(\tau_{0})\}=(2\pi)^{3}\delta^{ss^{\prime}}\delta^{3}(\boldsymbol{k}-\boldsymbol{q}).
\end{align}Substituting \eqref{ModeChi} and \eqref{ModeConjugatePi} into \eqref{AntiCommutationDefine} gives the momentum-space normalization condition
\begin{align}
&\delta_{\mathrm{I}_{1}}^{\mathrm{J}_{1}}=\sum_{s}\big(x_{\mathrm{I}_{1},\boldsymbol{k}}^{(s)}(\tau)x_{\dot{\mathrm{I}}_{2},\boldsymbol{k}}^{(s)\dagger}(\tau)(\bar{\sigma}^{0})^{\dot{\mathrm{I}}_{2}\mathrm{J}_{1}}+y_{\mathrm{I}_{1},-\boldsymbol{k}}^{(s)}(\tau)y_{\dot{\mathrm{I}}_{2},-\boldsymbol{k}}^{(s)\dagger}(\tau)(\bar{\sigma}^{0})^{\dot{\mathrm{I}}_{2}\mathrm{J}_{1}}\big).
\end{align}Using the helicity decomposition \eqref{EigenSpinorDecomModev1}--\eqref{EigenSpinorDecomModev2}, this relation becomes
\begin{align}
\label{NormalEigenSpinor}
&\delta_{\mathrm{I}_{1}}^{\mathrm{J}_{1}}=\sum_{s}\big(\vert X_{k}^{(s)}(\tau)\vert^{2}\xi_{s,\mathrm{I}_{1}}(\tilde{\boldsymbol{k}})\xi_{s,\dot{\mathrm{I}}_{2}}^{\dagger}(\tilde{\boldsymbol{k}})+\vert Y_{k}^{(s)}(\tau)\vert^{2}\xi_{s,\mathrm{I}_{1}}^{\dagger}(-\tilde{\boldsymbol{k}})\xi_{s,\dot{\mathrm{I}}_{2}}(-\tilde{\boldsymbol{k}})\big)(\bar{\sigma}^{0})^{\dot{\mathrm{I}}_{2}\mathrm{J}_{1}}.
\end{align}In this work we focus on the small-$\vartheta$ regime, which suppresses the effects of the cubic interaction $\partial_\mu \delta\phi \tilde{\chi}^\dagger\bar{\sigma}^\mu\tilde{\chi}$ generated by inflaton fluctuations around the homogeneous background. In this limit, the helicity sectors obey the relations $\vert X_{k}^{(+\frac{1}{2})}(\tau)\vert^{2}=\vert Y_{k}^{(-\frac{1}{2})}(\tau)\vert^{2}$ and $\vert Y_{k}^{(+\frac{1}{2})}(\tau)\vert^{2}=\vert X_{k}^{(-\frac{1}{2})}(\tau)\vert^{2}$. Together with the helicity sums \eqref{EigenSpinorHelicitySumv3}--\eqref{EigenSpinorHelicitySumv4}, \eqref{NormalEigenSpinor} reduces to the simple Wronskian normalization
\begin{align}
\label{NormalizaInXandY}
&1=\vert X_{k}^{(+\frac{1}{2})}(\tau)\vert^{2}+\vert Y_{k}^{(+\frac{1}{2})}(\tau)\vert^{2}=\vert Y_{k}^{(-\frac{1}{2})}(\tau)\vert^{2}+\vert X_{k}^{(-\frac{1}{2})}(\tau)\vert^{2}.
\end{align}Finally, the far-past asymptotic expansion of the mode functions \eqref{SolXkPlus}--\eqref{SolYkMinus} at $k\tau\to-\infty$ is consistent with this normalization and gives a vanishing initial particle number, in agreement with the Bunch--Davies vacuum condition and with \eqref{ParticleNumberFromBogoliubov}.

\section{Quadratic Hamiltonian and Bogoliubov transformation \label{AppendixC}}

For later use, we rewrite the mode expansions of the Majorana field and its
conjugate momentum by grouping the operators associated with each momentum
pair $(\boldsymbol{k},-\boldsymbol{k})$.  Using the helicity-spinor
decomposition introduced in Appendix~\ref{AppendixB}, the field operator and its Hermitian
conjugate take the forms given in Eqs.~\eqref{RewriteChiDownI1} and
\eqref{RewriteChiDaggerDownIdot1}, respectively.  The corresponding
conjugate-momentum operators are given in Eqs.~\eqref{RewritePiUpJ1} and
\eqref{RewritePiDaggerUpJdot1}.
\begin{small}
\begin{align}
\nonumber
\hat{\chi}_{\mathrm{I}_{1}}(\tau,\boldsymbol{x})&=\int\frac{d^{3}\boldsymbol{k}}{(2\pi)^{3}}\underbrace{\sum_{s=\pm\frac{1}{2}}\big(x_{\mathrm{I}_{1},\boldsymbol{k}}^{(s)}(\tau)\hat{a}_{\boldsymbol{k}}^{(s)}(\tau_{0})+y_{\mathrm{I}_{1},-\boldsymbol{k}}^{(s)}(\tau)\hat{a}_{-\boldsymbol{k}}^{(s)\dagger}(\tau_{0})\big)}_{\hat{\chi}_{\mathrm{I}_{1},\boldsymbol{k}}(\tau)}\text{e}^{\text{i}\boldsymbol{k}\cdot\boldsymbol{x}}\\
\label{RewriteChiDownI1}
&=\int\frac{d^{3}\boldsymbol{k}}{(2\pi)^{3}}\underbrace{\sum_{s=\pm\frac{1}{2}}\big(X_{k}^{(s)}(\tau)\xi_{s,\mathrm{I}_{1}}(\tilde{\boldsymbol{k}})\hat{a}_{\boldsymbol{k}}^{(s)}(\tau_{0})+Y_{k}^{(s)}(\tau)\xi_{s,\mathrm{I}_{1}}^{\dagger}(-\tilde{\boldsymbol{k}})\hat{a}_{-\boldsymbol{k}}^{(s)\dagger}(\tau_{0})\big)}_{\hat{\chi}_{\mathrm{I}_{1},\boldsymbol{k}}(\tau)}\text{e}^{\text{i}\boldsymbol{k}\cdot\boldsymbol{x}},\\
\nonumber
\hat{\chi}_{\dot{\mathrm{I}}_{1}}^{\dagger}(\tau,\boldsymbol{x})&=\int\frac{d^{3}\boldsymbol{k}}{(2\pi)^{3}}\underbrace{\sum_{s=\pm\frac{1}{2}}\big(x_{\dot{\mathrm{I}}_{1},\boldsymbol{k}}^{(s)\dagger}(\tau)\hat{a}_{\boldsymbol{k}}^{(s)\dagger}(\tau_{0})+y_{\dot{\mathrm{I}}_{1},-\boldsymbol{k}}^{(s)\dagger}(\tau)\hat{a}_{-\boldsymbol{k}}^{(s)}(\tau_{0})\big)}_{\hat{\chi}_{\dot{\mathrm{I}}_{1},\boldsymbol{k}}^{\dagger}(\tau)}\text{e}^{-\text{i}\boldsymbol{k}\cdot\boldsymbol{x}}\\
\label{RewriteChiDaggerDownIdot1}
&=\int\frac{d^{3}\boldsymbol{k}}{(2\pi)^{3}}\underbrace{\sum_{s=\pm\frac{1}{2}}\big(X_{k}^{(s)\star}(\tau)\xi_{s,\dot{\mathrm{I}}_{1}}^{\dagger}(\tilde{\boldsymbol{k}})\hat{a}_{\boldsymbol{k}}^{(s)\dagger}(\tau_{0})+Y_{k}^{\star(s)}(\tau)\xi_{s,\dot{\mathrm{I}}_{1}}(-\tilde{\boldsymbol{k}})\hat{a}_{-\boldsymbol{k}}^{(s)}(\tau_{0})\big)}_{\hat{\chi}_{\dot{\mathrm{I}}_{1},\boldsymbol{k}}^{\dagger}(\tau)}\text{e}^{-\text{i}\boldsymbol{k}\cdot\boldsymbol{x}},\\
\nonumber
\hat{\pi}_{(\chi)}^{\mathrm{J}_{1}}(\tau,\boldsymbol{x})&=\int\frac{d^{3}\boldsymbol{k}}{(2\pi)^{3}}\underbrace{\text{i}\sum_{s=\pm\frac{1}{2}}\big(x_{\dot{\mathrm{I}}_{2},\boldsymbol{k}}^{(s)\dagger}(\tau)\hat{a}_{\boldsymbol{k}}^{(s)\dagger}(\tau_{0})+y_{\dot{\mathrm{I}}_{2},-\boldsymbol{k}}^{(s)\dagger}(\tau)\hat{a}_{-\boldsymbol{k}}^{(s)}(\tau_{0})\big)(\bar{\sigma}^{0})^{\dot{\mathrm{I}}_{2}\mathrm{J}_{1}}}_{\hat{\pi}_{(\chi),-\boldsymbol{k}}^{\mathrm{J}_{1}}(\tau)}\text{e}^{-\text{i}\boldsymbol{k}\cdot\boldsymbol{x}}\\
\label{RewritePiUpJ1}
&=\int\frac{d^{3}\boldsymbol{k}}{(2\pi)^{3}}\underbrace{\text{i}\sum_{s=\pm\frac{1}{2}}\big(X_{k}^{(s)\star}(\tau)\xi_{s,\dot{\mathrm{I}}_{2}}^{\dagger}(\tilde{\boldsymbol{k}})\hat{a}_{\boldsymbol{k}}^{(s)\dagger}(\tau_{0})+Y_{k}^{\star(s)}(\tau)\xi_{s,\dot{\mathrm{I}}_{2}}(-\tilde{\boldsymbol{k}})\hat{a}_{-\boldsymbol{k}}^{(s)}(\tau_{0})\big)(\bar{\sigma}^{0})^{\dot{\mathrm{I}}_{2}\mathrm{J}_{1}}}_{\hat{\pi}_{(\chi),-\boldsymbol{k}}^{\mathrm{J}_{1}}(\tau)}\text{e}^{-\text{i}\boldsymbol{k}\cdot\boldsymbol{x}},\\
\nonumber
\hat{\pi}_{(\chi)}^{\dagger\dot{\mathrm{J}}_{1}}(\tau,\boldsymbol{x})&=\int\frac{d^{3}\boldsymbol{k}}{(2\pi)^{3}}\underbrace{-\text{i}\sum_{s=\pm\frac{1}{2}}(\bar{\sigma}^{0})^{\dot{\mathrm{J}}_{1}\mathrm{I}_{2}}\big(x_{\mathrm{I}_{2},\boldsymbol{k}}^{(s)}(\tau)\hat{a}_{\boldsymbol{k}}^{(s)}(\tau_{0})+y_{\mathrm{I}_{2},-\boldsymbol{k}}^{(s)}(\tau)\hat{a}_{-\boldsymbol{k}}^{(s)\dagger}(\tau_{0})\big)}_{\hat{\pi}_{(\chi),-\boldsymbol{k}}^{\dagger\dot{\mathrm{J}}_{1}}(\tau)}\text{e}^{\text{i}\boldsymbol{k}\cdot\boldsymbol{x}}\\
\label{RewritePiDaggerUpJdot1}
&=\int\frac{d^{3}\boldsymbol{k}}{(2\pi)^{3}}\underbrace{-\text{i}\sum_{s=\pm\frac{1}{2}}(\bar{\sigma}^{0})^{\dot{\mathrm{J}}_{1}\mathrm{I}_{2}}\big(X_{k}^{(s)}(\tau)\xi_{s,\mathrm{I}_{2}}(\tilde{\boldsymbol{k}})\hat{a}_{\boldsymbol{k}}^{(s)}(\tau_{0})+Y_{k}^{(s)}(\tau)\xi_{s,\mathrm{I}_{2}}^{\dagger}(-\tilde{\boldsymbol{k}})\hat{a}_{-\boldsymbol{k}}^{(s)\dagger}(\tau_{0})\big)}_{\hat{\pi}_{(\chi),-\boldsymbol{k}}^{\dagger\dot{\mathrm{J}}_{1}}(\tau)}\text{e}^{\text{i}\boldsymbol{k}\cdot\boldsymbol{x}}.
\end{align}
\end{small}In deriving Eqs.~\eqref{RewritePiUpJ1} and
\eqref{RewritePiDaggerUpJdot1}, we used
$\hat{\pi}_{(\chi),-\boldsymbol{k}}^{\mathrm{J}_{1}}
=\mathrm{i}\hat{\chi}_{\dot{\mathrm{I}}_{2},\boldsymbol{k}}^{\dagger}
(\bar{\sigma}^{0})^{\dot{\mathrm{I}}_{2}\mathrm{J}_{1}}$ and its Hermitian
conjugate.  The Hamiltonian density associated with the rescaled Majorana
action can then be written in the form\begin{align}
\nonumber
\mathcal{H}_{\text{(Majo)}}&=-\frac{\text{i}}{2}\chi_{\dot{\mathrm{I}}_{1}}^{\dagger}(\bar{\sigma}^{i})^{\dot{\mathrm{I}}_{1}\mathrm{J}_{1}}\partial_{i}\chi_{\mathrm{J}_{1}}-\frac{\text{i}}{2}\chi^{\mathrm{J}_{1}}(\sigma^{i})_{\mathrm{J}_{1}\dot{\mathrm{I}}_{1}}\partial_{i}\chi^{\dagger\dot{\mathrm{I}}_{1}}+\frac{a(\tau)m}{2}(\chi^{\mathrm{I}_{1}}\chi_{\mathrm{I}_{1}}+\chi_{\dot{\mathrm{I}}_{1}}^{\dagger}\chi^{\dagger,\dot{\mathrm{I}}_{1}})\\
\label{NormalHamilDensity}
&-\frac{C_{\text{A-F}}\bar{\phi}^{\prime}}{2f}\chi_{\dot{\mathrm{I}}_{1}}^{\dagger}(\bar{\sigma}^{0})^{\dot{\mathrm{I}}_{1}\mathrm{J}_{1}}\chi_{\mathrm{J}_{1}}+\frac{C_{\text{A-F}}\bar{\phi}^{\prime}}{2f}\chi^{\mathrm{J}_{1}}(\sigma^{0})_{\mathrm{J}_{1}\dot{\mathrm{I}}_{1}}\chi^{\dagger\dot{\mathrm{I}}_{1}}.
\end{align}After promoting the classical fields in Eq.~\eqref{NormalHamilDensity} to
operators and substituting Eqs.~\eqref{RewriteChiDownI1}--
\eqref{RewritePiDaggerUpJdot1}, we obtain the quadratic Hamiltonian operator
\begin{small}
\begin{align}
\nonumber
\hat{H}_{\text{(Majo)}}(\tau)&=\sum_{s=\pm\frac{1}{2}}\int\frac{d^{3}\boldsymbol{k}}{(2\pi)^{3}}\bigg(\underbrace{\big\{2(sk+\frac{C_{\text{A-F}}\bar{\phi}^{\prime}}{2f})X_{k}^{(s)}(\tau)Y_{k}^{\star(s)}(\tau)+\frac{ma}{2}\big(X_{k}^{(s)}(\tau)X_{k}^{(s)}(\tau)-Y_{k}^{\star(s)}(\tau)Y_{k}^{\star(s)}(\tau)\big)\big\}}_{\mathcal{B}_{k}^{(s)}(\tau)}\!\times\hat{a}_{-\boldsymbol{k}}^{(s)}(\tau_{0})\hat{a}_{\boldsymbol{k}}^{(s)}(\tau_{0})\\
\nonumber
&\hspace{-5mm}+\!\underbrace{\big\{2(sk\!+\frac{C_{\text{A-F}}\bar{\phi}^{\prime}}{2f})X_{k}^{(s)\star}(\tau)Y_{k}^{(s)}(\tau)\!+\frac{ma}{2}\big(X_{k}^{(s)\star}(\tau)X_{k}^{(s)\star}(\tau)\!-\!Y_{k}^{(s)}(\tau)Y_{k}^{(s)}(\tau)\big)\big\}}_{\mathcal{B}_{k}^{(s)\star}(\tau)}\!\times\hat{a}_{\boldsymbol{k}}^{(s)\dagger}(\tau_{0})\hat{a}_{-\boldsymbol{k}}^{(s)\dagger}(\tau_{0})\\
\nonumber
&\hspace{-5mm}+\!\underbrace{\big\{(sk\!+\frac{C_{\text{A-F}}\bar{\phi}^{\prime}}{2f})\big(\vert X_{k}^{(s)}(\tau)\vert^{2}\!-\!\vert Y_{k}^{(s)}(\tau)\vert^{2}\big)\!-\frac{ma}{2}\big(X_{k}^{(s)}(\tau)Y_{k}^{(s)}(\tau)\!+\!X_{k}^{(s)\star}(\tau)Y_{k}^{\star(s)}(\tau)\big)\big\}}_{-\mathcal{A}_{k}^{(s)}(\tau)}\!\times\hat{a}_{-\boldsymbol{k}}^{(s)}(\tau_{0})\hat{a}_{-\boldsymbol{k}}^{(s)\dagger}(\tau_{0})\\
\label{QuarticHamiltonMatrix}
&\hspace{-5mm}+\underbrace{\big\{(sk\!+\frac{C_{\text{A-F}}\bar{\phi}^{\prime}}{2f})\big(\vert Y_{k}^{(s)}(\tau)\vert^{2}\!-\!\vert X_{k}^{(s)}(\tau)\vert^{2}\big)+\frac{ma}{2}\big(X_{k}^{(s)}(\tau)Y_{k}^{(s)}(\tau)\!+\!X_{k}^{(s)\star}(\tau)Y_{k}^{\star(s)}(\tau)\big)\big\}}_{\mathcal{A}_{k}^{(s)}(\tau)~,~\text{note that }\mathcal{A}_{k}^{(s)\star}(\tau)=\mathcal{A}_{k}^{(s)}(\tau)}\!\times\hat{a}_{\boldsymbol{k}}^{(s)\dagger}(\tau_{0})\hat{a}_{\boldsymbol{k}}^{(s)}(\tau_{0})\bigg).
\end{align}
\end{small}The coefficient $\mathcal A_k^{(s)}(\tau)$ is real, whereas
$\mathcal B_k^{(s)}(\tau)$ is generally complex.  The terms proportional to
$\mathcal A_k^{(s)}$ are diagonal in the initial particle basis.  By contrast,
$\mathcal B_k^{(s)}$ and $\mathcal B_k^{(s)\star}$ multiply the pair-annihilation
and pair-creation operators, respectively, and therefore encode the
nonadiabatic mixing responsible for Majorana-pair production. Introducing the Nambu basis
$(\hat a_{\boldsymbol{k}}^{(s)},
\hat a_{-\boldsymbol{k}}^{(s)\dagger})^{\mathrm T}$, the Hamiltonian in
Eq.~\eqref{QuarticHamiltonMatrix} can be written compactly as
\begin{align}
\label{Hamilto}
&\hat{H}_{\text{(Majo)}}(\tau)=\sum_{s=\pm\frac{1}{2}}\int\frac{d^{3}\boldsymbol{k}}{(2\pi)^{3}}\left(\begin{array}{cc}
\hat{a}_{\boldsymbol{k}}^{(s)\dagger}(\tau_{0}) & \hat{a}_{-\boldsymbol{k}}^{(s)}(\tau_{0})\end{array}\right)\underbrace{\left(\begin{array}{cc}
	\mathcal{A}_{k}^{(s)}(\tau) & \mathcal{B}_{k}^{(s)\star}(\tau)\\
	\mathcal{B}_{k}^{(s)}(\tau) & -\mathcal{A}_{k}^{(s)}(\tau)
\end{array}\right)}_{\boldsymbol{\mathcal{M}}_{k}^{(s)}(\tau)}\left(\begin{array}{c}
\hat{a}_{\boldsymbol{k}}^{(s)}(\tau_{0})\\
\hat{a}_{-\boldsymbol{k}}^{(s)\dagger}(\tau_{0})
\end{array}\right).
\end{align}The matrix $\boldsymbol{\mathcal M}_k^{(s)}(\tau)$ in Eq.~\eqref{Hamilto} is
Hermitian and traceless.  Its eigenvalues are therefore
$\pm\omega_k^{(s)}(\tau)$.  It is diagonalized by the unitary matrix
$\boldsymbol{\mathcal U}_k^{(s)}(\tau)$ according to
\begin{align}
\label{DiagnolUnitaryDecom}
&\boldsymbol{\mathcal{M}}_{k}^{(s)}(\tau)=\boldsymbol{\mathcal{U}}_{k}^{(s)\dagger}(\tau)\cdot\left(\begin{array}{cc}
\omega_{k}^{(s)}(\tau) & 0\\
0 & -\omega_{k}^{(s)}(\tau)
\end{array}\right)\cdot\boldsymbol{\mathcal{U}}_{k}^{(s)}(\tau),\\
\nonumber
&\boldsymbol{\mathcal{U}}_{k}^{(s)}(\tau)=\frac{1}{\sqrt{2\omega_{k}^{(s)}}}\left(\begin{array}{cc}
\frac{\mathcal{B}_{k}^{(s)}}{\sqrt{\omega_{k}^{(s)}-\mathcal{A}_{k}^{(s)}}} & \sqrt{\omega_{k}^{(s)}-\mathcal{A}_{k}^{(s)}}\\
\frac{\mathcal{B}_{k}^{(s)}}{\sqrt{\omega_{k}^{(s)}+\mathcal{A}_{k}^{(s)}}} & -\sqrt{\omega_{k}^{(s)}+\mathcal{A}_{k}^{(s)}}
\end{array}\right),\\
\nonumber
&\omega_{k}^{(s)}(\tau)=\sqrt{(sk+\frac{\vartheta}{2\tau})^{2}+\frac{m^{2}a^{2}}{4}}.
\end{align}In the last line of Eq.~\eqref{DiagnolUnitaryDecom}, we used
Eq.~\eqref{AxionCouplingWithSRone} to replace
$C_{\mathrm{A-F}}\bar\phi'/f$ by $aH\vartheta$.  The diagonalization defines
the instantaneous quasiparticle basis and gives the Bogoliubov transformation
between the operators at time $\tau$ and those defined with respect to the
Bunch--Davies vacuum at $\tau_0$:
\begin{align}
\label{StandardBogoTransCreatAnnihi}
&\left(\begin{array}{c}
\hat{a}_{\boldsymbol{k}}^{(s)}(\tau)\\
\hat{a}_{-\boldsymbol{k}}^{(s)\dagger}(\tau)
\end{array}\right)=\left(\begin{array}{cc}
\frac{\mathcal{B}_{k}^{(s)}}{\sqrt{2\omega_{k}^{(s)}}\sqrt{\omega_{k}^{(s)}-\mathcal{A}_{k}^{(s)}}} & \frac{\sqrt{\omega_{k}^{(s)}-\mathcal{A}_{k}^{(s)}}}{\sqrt{2\omega_{k}^{(s)}}}\\
\frac{1}{\sqrt{2\omega_{k}^{(s)}}}\frac{\mathcal{B}_{k}^{(s)}}{\sqrt{\omega_{k}^{(s)}+\mathcal{A}_{k}^{(s)}}} & -\frac{\sqrt{\omega_{k}^{(s)}+\mathcal{A}_{k}^{(s)}}}{\sqrt{2\omega_{k}^{(s)}}}
\end{array}\right)\left(\begin{array}{c}
\hat{a}_{\boldsymbol{k}}^{(s)}(\tau_{0})\\
\hat{a}_{-\boldsymbol{k}}^{(s)\dagger}(\tau_{0})
\end{array}\right).
\end{align}Because $\boldsymbol{\mathcal U}_k^{(s)}$ is unitary, the transformation in
Eq.~\eqref{StandardBogoTransCreatAnnihi} preserves the canonical
anticommutation relations. Its off-diagonal entries mix
$\hat a_{\boldsymbol{k}}^{(s)}(\tau_0)$ with
$\hat a_{-\boldsymbol{k}}^{(s)\dagger}(\tau_0)$ and thus describe fermionic
pair production. For a mode $(\boldsymbol{k},s)$, the instantaneous occupation number in the
Bunch--Davies vacuum is most directly defined by
\begin{align}
\nonumber
\mathcal{N}_{k}^{(s)}(\tau)&=\sum_{s^{\prime}}\int_{\mathbb{R}^{3+}}\frac{d^{3}\boldsymbol{q}}{(2\pi)^{3}}\langle\Psi(\tau)\vert\hat{a}_{\boldsymbol{k}}^{(s)}(\tau_{0})\hat{a}_{\boldsymbol{q}}^{(s^{\prime})}(\tau_{0})\vert\Psi(\tau)\rangle\\
\nonumber
&=\!\sum_{s^{\prime}}\int_{\mathbb{R}^{3+}}\frac{d^{3}\boldsymbol{q}}{(2\pi)^{3}}\,_{\tau_{0}}\!\langle0\vert\hat{\mathcal{U}}^{-1}(\tau,\tau_{0})\hat{a}_{\boldsymbol{k}}^{(s)}(\tau_{0})\hat{\mathcal{U}}(\tau,\tau_{0})\cdot\hat{\mathcal{U}}^{-1}(\tau,\tau_{0})\hat{a}_{\boldsymbol{q}}^{(s^{\prime})}(\tau_{0})\hat{\mathcal{U}}(\tau,\tau_{0})\vert0\rangle_{\tau_{0}}\\
\label{DefineTheParticleNum}
&=\sum_{s^{\prime}}\int_{\mathbb{R}^{3+}}\frac{d^{3}\boldsymbol{q}}{(2\pi)^{3}}\,_{\tau_{0}}\!\langle0\vert\hat{a}_{\boldsymbol{k}}^{(s)}(\tau)\hat{a}_{\boldsymbol{q}}^{(s^{\prime})}(\tau)\vert0\rangle_{\tau_{0}}.
\end{align}Substituting Eq.~\eqref{StandardBogoTransCreatAnnihi} into
Eq.~\eqref{DefineTheParticleNum} gives
\begin{align}
\label{ParticleNumberFromBogoliubov}
&\mathcal{N}_{k}^{(s)}=\frac{1}{2\omega_{k}^{(s)}}\frac{\vert\mathcal{B}_{k}^{(s)}\vert^{2}}{\omega_{k}^{(s)}+\mathcal{A}_{k}^{(s)}}=\frac{1}{2\omega_{k}^{(s)}}\frac{\omega_{k}^{(s)2}-\mathcal{A}_{k}^{(s)2}}{\omega_{k}^{(s)}+\mathcal{A}_{k}^{(s)}}=\frac{\omega_{k}^{(s)}-\mathcal{A}_{k}^{(s)}}{2\omega_{k}^{(s)}}.
\end{align}At the initial time $\tau_0$, the Bunch--Davies condition
$\mathcal N_k^{(s)}(\tau_0)=0$, together with
Eq.~\eqref{ParticleNumberFromBogoliubov}, implies
$\mathcal B_k^{(s)}(\tau_0)=0$, or equivalently
$\operatorname{Re}\mathcal B_k^{(s)}(\tau_0)
=\operatorname{Im}\mathcal B_k^{(s)}(\tau_0)=0$.  This result can also be
read directly from the structure of the Hamiltonian in
Eq.~\eqref{QuarticHamiltonMatrix}.  Indeed, $\mathcal B_k^{(s)}$ and
$\mathcal B_k^{(s)\star}$ are precisely the coefficients of the
pair-annihilation and pair-creation operators.  Their vanishing at $\tau_0$
therefore removes the off-diagonal terms that mix the vacuum with the
two-particle state, in agreement with the absence of particles in the initial
state.  The same conclusion becomes particularly transparent after rewriting
the Hamiltonian in terms of the $su(2)$ generators
\eqref{OpeJplus}--\eqref{OpeJz}.  As shown in
Eq.~\eqref{RewriteHamiltonian}, the coefficients of the ladder, or
weight-shifting, generators
$\hat{\mathcal J}_{+,\boldsymbol{k}}^{(s)}$ and
$\hat{\mathcal J}_{-,\boldsymbol{k}}^{(s)}$ are proportional to
$\mathcal B_k^{(s)\star}$ and $\mathcal B_k^{(s)}$, respectively.  When
$\mathcal B_k^{(s)}(\tau_0)=0$, only the diagonal
$\hat{\mathcal J}_{z,\boldsymbol{k}}^{(s)}$ contribution remains; this term
does not change the occupation number and hence cannot produce particle
pairs.  At later times, the evolving background can generate a nonzero
$\mathcal B_k^{(s)}(\tau)$, activate the ladder-generator terms, and thereby
produce correlated Majorana pairs.

\section{Bogoliubov transformation, time evolution, and the fermionic
two-mode state \label{SqueeTwoModeBuild}}

In this appendix, we derive the time-dependent Bogoliubov transformation in
the fermionic squeezing formalism and construct the corresponding two-mode
state.  Our goal is to relate the operators
$\hat a_{\pm\boldsymbol{k}}^{(s)}(\tau)$ and
$\hat a_{\pm\boldsymbol{k}}^{(s)\dagger}(\tau)$ at an arbitrary time $\tau$
to the operators defined with respect to the Bunch--Davies vacuum at
$\tau_0$.

The quadratic Hamiltonian acts independently in each helicity and momentum
pair $(\boldsymbol{k},-\boldsymbol{k})$.  To expose its algebraic structure,
we introduce the pair-creation, pair-annihilation, and number generators in
Eqs.~\eqref{OpeJplus}--\eqref{OpeJz}.
\begin{align}
\label{OpeJplus}
&\hat{\mathcal{J}}_{+}(\tau_{0})=\sum_{s}\int_{\mathbb{R}^{3+}}\frac{d^{3}\boldsymbol{k}}{(2\pi)^{3}}\hat{\mathcal{J}}_{+,\boldsymbol{k}}^{(s)}(\tau_{0})=\sum_{s}\int_{\mathbb{R}^{3+}}\frac{d^{3}\boldsymbol{k}}{(2\pi)^{3}}\hat{a}_{-\boldsymbol{k}}^{(s)\dagger}(\tau_{0})\hat{a}_{\boldsymbol{k}}^{(s)\dagger}(\tau_{0}),\\
\label{OpeJminus}
&\hat{\mathcal{J}}_{-}(\tau_{0})=\sum_{s}\int_{\mathbb{R}^{3+}}\frac{d^{3}\boldsymbol{k}}{(2\pi)^{3}}\hat{\mathcal{J}}_{-,\boldsymbol{k}}^{(s)}(\tau_{0})=\sum_{s}\int_{\mathbb{R}^{3+}}\frac{d^{3}\boldsymbol{k}}{(2\pi)^{3}}\hat{a}_{\boldsymbol{k}}^{(s)}(\tau_{0})\hat{a}_{-\boldsymbol{k}}^{(s)}(\tau_{0}),\\
\label{OpeJz}
&\hat{\mathcal{J}}_{z}(\tau_{0})\!=\!\sum_{s}\int_{\mathbb{R}^{3+}}\frac{d^{3}\boldsymbol{k}}{(2\pi)^{3}}\!\hat{\mathcal{J}}_{z,\boldsymbol{k}}^{(s)}(\tau_{0})\!=\!\sum_{s}\int_{\mathbb{R}^{3+}}\frac{d^{3}\boldsymbol{k}}{(2\pi)^{3}}\frac{1}{2}\big(\hat{a}_{-\boldsymbol{k}}^{(s)\dagger}(\tau_{0})\hat{a}_{-\boldsymbol{k}}^{(s)}(\tau_{0})\!-\!\hat{a}_{\boldsymbol{k}}^{(s)}(\tau_{0})\hat{a}_{\boldsymbol{k}}^{(s)\dagger}(\tau_{0})\big).
\end{align}The integration domain $\mathbb R^{3+}$ denotes one half of momentum space;
this restriction avoids double counting the paired modes
$(\boldsymbol{k},-\boldsymbol{k})$. From these operators, one can readily verify the following commutation relations, which constitute the $su(2)$ Lie algebra:
\begin{align}
\label{Closedsu2CreatAnnihilate}
&[\hat{\mathcal{J}}_{+}(\tau_0),\hat{\mathcal{J}}_{-}(\tau_0)]=2\hat{\mathcal{J}}_{z}(\tau_0)~,~[\hat{\mathcal{J}}_{z}(\tau_{0}),\hat{\mathcal{J}}_{+}(\tau_{0})]=\hat{\mathcal{J}}_{+}(\tau_{0})~,~[\hat{\mathcal{J}}_{z}(\tau_{0}),\hat{\mathcal{J}}_{-}(\tau_{0})]=-\hat{\mathcal{J}}_{-}(\tau_{0}).
\end{align}In deriving \eqref{Closedsu2CreatAnnihilate} from the operator definitions given in \eqref{OpeJplus}--\eqref{OpeJz}, it is necessary to employ the following identities for the nested commutators involving Grassmann-valued operators:
\begin{align}
\nonumber
[\hat{\mathcal{O}}_{1}^{\text{(Grass)}}\hat{\mathcal{O}}_{2}^{\text{(Grass)}},\hat{\mathcal{O}}_{3}^{\text{(Grass)}}\hat{\mathcal{O}}_{4}^{\text{(Grass)}}]&=\hat{\mathcal{O}}_{1}^{\text{(Grass)}}\{\hat{\mathcal{O}}_{2}^{\text{(Grass)}},\hat{\mathcal{O}}_{3}^{\text{(Grass)}}\}\hat{\mathcal{O}}_{4}^{\text{(Grass)}}-\hat{\mathcal{O}}_{1}^{\text{(Grass)}}\hat{\mathcal{O}}_{3}^{\text{(Grass)}}\{\hat{\mathcal{O}}_{2}^{\text{(Grass)}},\hat{\mathcal{O}}_{4}^{\text{(Grass)}}\}\\
&-\hat{\mathcal{O}}_{3}^{\text{(Grass)}}\{\hat{\mathcal{O}}_{1}^{\text{(Grass)}},\hat{\mathcal{O}}_{4}^{\text{(Grass)}}\}\hat{\mathcal{O}}_{2}^{\text{(Grass)}}\!+\!\{\hat{\mathcal{O}}_{1}^{\text{(Grass)}},\hat{\mathcal{O}}_{3}^{\text{(Grass)}}\}\hat{\mathcal{O}}_{4}^{\text{(Grass)}}\hat{\mathcal{O}}_{2}^{\text{(Grass)}}.
\end{align}Using Eqs.~\eqref{OpeJplus}--\eqref{OpeJz}, the Hamiltonian in
Eq.~\eqref{QuarticHamiltonMatrix} becomes
\begin{align}
\label{RewriteHamiltonian}
&\hat{H}_{\text{(Majo)}}(\tau)=\sum_{s=\pm\frac{1}{2}}\int_{\mathbb{R}^{3+}}\frac{d^{3}\boldsymbol{k}}{(2\pi)^{3}}\big\{-\mathcal{B}_{k}^{(s)\star}\hat{\mathcal{J}}_{+,\boldsymbol{k}}^{(s)}(\tau_{0})-\mathcal{B}_{k}^{(s)}\hat{\mathcal{J}}_{-,\boldsymbol{k}}^{(s)}(\tau_{0})+2\mathcal{A}_{k}^{(s)}\hat{\mathcal{J}}_{z,\boldsymbol{k}}^{(s)}(\tau_{0})\big\}.
\end{align}Equation~\eqref{RewriteHamiltonian} separates the diagonal contribution,
proportional to
$\mathcal A_k^{(s)}\hat{\mathcal J}_{z,\boldsymbol{k}}^{(s)}$, from the
off-diagonal pairing contributions proportional to
$\mathcal B_k^{(s)}\hat{\mathcal J}_{-,\boldsymbol{k}}^{(s)}$ and its complex
conjugate.  The latter mix the vacuum and one-pair sectors. With the standard Schrodinger convention, the time evolution operator is
\begin{align}
\label{TimeEvolutionOperator}
\hat{\mathcal U}(\tau,\tau_0)
&=\mathcal T\exp\!\left[-\mathrm{i}
\int_{\tau_0}^{\tau}d\tau'\,
\hat H_{\mathrm{(Majo)}}(\tau')\right].
\end{align}A direct Zassenhaus expansion  of Eq.~\eqref{TimeEvolutionOperator} generates
an infinite sequence of nested commutators \cite{Casas:2012nqk,Kimura:2017xxz}.  This is unnecessary here because
$\hat H_{\mathrm{(Majo)}}$ is a linear combination of $su(2)$ generators, so
the time-ordered exponential remains an element of the corresponding
$SU(2)$ group.  It can therefore be parameterized as a fermionic squeezing
operator followed by a number rotation:
\begin{align}
\nonumber
\hat{\mathcal{U}}(\tau,\tau_{0})&=\!\underbrace{\exp\bigg(\text{i}\sum_{s}\int_{\mathbb{R}^{3+}}\frac{d^{3}\boldsymbol{k}}{(2\pi)^{3}}r_{k}^{(s)}(\tau)\big\{\text{e}^{-\text{i}\varphi_{k}^{(s)}(\tau)}\hat{\mathcal{J}}_{+,\boldsymbol{k}}^{(s)}(\tau_{0})+\text{e}^{\text{i}\varphi_{k}^{(s)}(\tau)}\hat{\mathcal{J}}_{-,\boldsymbol{k}}^{(s)}(\tau_{0})\big\}\bigg)}_{\hat{\mathcal{S}}(r,\varphi)}\\
\label{TimeEvoSU2Symmetry}
&\cdot \underbrace{\exp\bigg(2\text{i}\sum_{s}\int_{\mathbb{R}^{3+}}\frac{d^{3}\boldsymbol{k}}{(2\pi)^{3}}\omega_{k}^{(s)}(\tau)\hat{\mathcal{J}}_{z,\boldsymbol{k}}^{(s)}(\tau_{0})\bigg)}_{\hat{\mathcal{R}}(\omega)}.
\end{align}The time-dependent parameters $r_k^{(s)}(\tau)$,
$\varphi_k^{(s)}(\tau)$, and $\omega_k^{(s)}(\tau)$ are not arbitrary.  They
are fixed by the Schrodinger equation, or equivalently by requiring the
operator transformation generated by Eq.~\eqref{TimeEvoSU2Symmetry} to agree
with Eq.~\eqref{StandardBogoTransCreatAnnihi}.  Accordingly,
Eq.~\eqref{TimeEvoSU2Symmetry} should be understood as an exact $SU(2)$
parameterization of the time-ordered evolution, rather than as an
approximation in which time ordering is discarded. For generators obeying Eq.~\eqref{Closedsu2CreatAnnihilate}, the standard
$SU(2)$ disentangling identity is \cite{Puri:2001,Barnett},
\begin{align}
\label{GeneSU2Decomposition}
&\exp\big(\text{i}\theta_{+}\hat{\mathcal{J}}_{+}+\text{i}\theta_{-}\hat{\mathcal{J}}_{-}+\text{i}\theta_{3}\hat{\mathcal{J}}_{z}\big)=\exp\big(\vartheta_{+}\hat{\mathcal{J}}_{+}\big)\exp\big(\vartheta_{3}\hat{\mathcal{J}}_{z}\big)\exp\big(\vartheta_{-}\hat{\mathcal{J}}_{-}\big).\\
\nonumber
&\Gamma_{1}^{2}=\theta_{+}\theta_{-}+\frac{\theta_{3}^{2}}{4}~,~\vartheta_{3}=-2\ln\big(\cos(\Gamma_{1})-\frac{\text{i}\theta_{3}}{2\Gamma_{1}}\sin(\Gamma_{1})\big),\\
\nonumber
&\vartheta_{+}=\frac{\text{i}\theta_{+}}{\Gamma_{1}}\frac{\sin(\Gamma_{1})}{\cos(\Gamma_{1})-\text{i}\theta_{3}\sin(\Gamma_{1})/2\Gamma_{1}}~,~\vartheta_{-}=\frac{\text{i}\theta_{-}}{\Gamma_{1}}\frac{\sin(\Gamma_{1})}{\cos(\Gamma_{1})-\text{i}\theta_{3}\sin(\Gamma_{1})/2\Gamma_{1}}.
\end{align}Applying Eq.~\eqref{GeneSU2Decomposition} to the $\hat{\mathcal{S}}(r,\varphi)$ part of
Eq.~\eqref{TimeEvoSU2Symmetry} gives the normally ordered expression
\begin{align}
\nonumber
\hat{\mathcal{S}}(r,\varphi)&=\underbrace{\exp\bigg(\text{i}\sum_{s}\int_{\mathbb{R}^{3+}}\frac{d^{3}\boldsymbol{k}}{(2\pi)^{3}}\text{e}^{-\text{i}\varphi_{k}^{(s)}(\tau)}\tan\big(r_{k}^{(s)}(\tau)\big)\hat{a}_{-\boldsymbol{k}}^{(s)\dagger}(\tau_{0})\hat{a}_{\boldsymbol{k}}^{(s)\dagger}(\tau_{0})\bigg)}_{\hat{\mathcal{T}}_{4}}\\
\nonumber
&\cdot\underbrace{\exp\bigg(\!-\sum_{s}\int_{\mathbb{R}^{3+}}\frac{d^{3}\boldsymbol{k}}{(2\pi)^{3}}\ln\big(\cos\big(r_{k}^{(s)}(\tau)\big)\big)\big(\hat{a}_{-\boldsymbol{k}}^{(s)\dagger}(\tau_{0})\hat{a}_{-\boldsymbol{k}}^{(s)}(\tau_{0})\!-\!\hat{a}_{\boldsymbol{k}}^{(s)}(\tau_{0})\hat{a}_{\boldsymbol{k}}^{(s)\dagger}(\tau_{0})\big)\bigg)}_{\hat{\mathcal{T}}_{3}}\\
\label{DecomOPEHatSCreaAnnihi}
&\cdot\underbrace{\exp\bigg(\text{i}\sum_{s}\int_{\mathbb{R}^{3+}}\frac{d^{3}\boldsymbol{k}}{(2\pi)^{3}}\text{e}^{\text{i}\varphi_{k}^{(s)}(\tau)}\tan\big(r_{k}^{(s)}(\tau)\big)\hat{a}_{\boldsymbol{k}}^{(s)}(\tau_{0})\hat{a}_{-\boldsymbol{k}}^{(s)}(\tau_{0})\bigg)}_{\hat{\mathcal{T}}_{2}}.
\end{align}The three factors in Eq.~\eqref{DecomOPEHatSCreaAnnihi} have distinct roles:
$\hat{\mathcal T}_2$ contains the pair-annihilation operator,
$\hat{\mathcal T}_3$ fixes the normalization, and
$\hat{\mathcal T}_4$ creates a correlated pair with momenta
$(\boldsymbol{k},-\boldsymbol{k})$.  The number-rotation operator can be
written as
\begin{align}
\label{OPEHatRCreaAnnihi}
&\hat{\mathcal{R}}(\omega)=\underbrace{\exp\bigg(\text{i}\sum_{s}\int_{\mathbb{R}^{3+}}\frac{d^{3}\boldsymbol{k}}{(2\pi)^{3}}\omega_{k}^{(s)}(\tau)\big(\hat{a}_{-\boldsymbol{k}}^{(s)\dagger}(\tau_{0})\hat{a}_{-\boldsymbol{k}}^{(s)}(\tau_{0})\!-\!\hat{a}_{\boldsymbol{k}}^{(s)}(\tau_{0})\hat{a}_{\boldsymbol{k}}^{(s)\dagger}(\tau_{0})\big)\bigg)}_{\hat{\mathcal{T}}_{1}}.
\end{align}Acting with Eqs.~\eqref{DecomOPEHatSCreaAnnihi} and
\eqref{OPEHatRCreaAnnihi} on the Bunch--Davies vacuum yields
\begin{align}
\nonumber
\vert\Psi_{r_{q}^{(s)},\varphi_{q}^{(s)},\omega_{q}^{(s)}}(\tau)\rangle&=\hat{\mathcal{U}}(\tau,\tau_{0})\vert0_{\boldsymbol{q}}^{(s)},0_{-\boldsymbol{q}}^{(s)}\rangle_{\tau_{0}}=\cos\big(r_{q}^{(s)}(\tau)\big)\text{e}^{-\text{i}\omega_{q}^{(s)}(\tau)}\vert0_{\boldsymbol{q}}^{(s)},0_{-\boldsymbol{q}}^{(s)}\rangle_{\tau_{0}}\\
\label{Task1QuanState}
&+\text{i}\sin\big(r_{q}^{(s)}(\tau)\big)\text{e}^{-\text{i}\big(\varphi_{q}^{(s)}(\tau)+\omega_{q}^{(s)}(\tau)\big)}\vert1_{\boldsymbol{q}}^{(s)},1_{-\boldsymbol{q}}^{(s)}\rangle_{\tau_{0}}.
\end{align}Unlike a bosonic squeezed state, the fermionic state in
Eq.~\eqref{Task1QuanState} contains only the vacuum and one-pair sectors.  This
finite truncation follows from the canonical anticommutation relations and
the nilpotency
$(\hat a_{\pm\boldsymbol{k}}^{(s)\dagger})^2=0$.  Equivalently, the Pauli
exclusion principle restricts the occupation number of each mode to either
$0$ or $1$.  The squeezing amplitude therefore determines a bounded
occupation number, $\mathcal N_k^{(s)}=\sin^2r_k^{(s)}$. We next derive the induced transformation of the annihilation operator.
Using
$\hat{\mathcal U}=\hat{\mathcal T}_4\hat{\mathcal T}_3
\hat{\mathcal T}_2\hat{\mathcal T}_1$ and commuting
$\hat a_{\boldsymbol{k}}^{(s)}(\tau_0)$ successively through these factors,
we obtain
\begin{align}
\nonumber
\hat{a}_{\boldsymbol{k}}^{(s)}(\tau)&=\!\hat{\mathcal{R}}^{-1}(\omega)\hat{\mathcal{S}}^{-1}(r,\varphi)\hat{a}_{\boldsymbol{k}}^{(s)}(\tau_{0})\hat{\mathcal{S}}(r,\varphi)\hat{\mathcal{R}}(\omega)\!=\!\hat{\mathcal{T}}_{1}^{-1}\hat{\mathcal{T}}_{2}^{-1}\hat{\mathcal{T}}_{3}^{-1}\hat{\mathcal{T}}_{4}^{-1}\cdot\hat{a}_{\boldsymbol{k}}^{(s)}(\tau_{0})\cdot\hat{\mathcal{T}}_{4}\hat{\mathcal{T}}_{3}\hat{\mathcal{T}}_{2}\hat{\mathcal{T}}_{1}\\
\nonumber
&=\hat{\mathcal{T}}_{1}^{-1}\hat{\mathcal{T}}_{2}^{-1}\hat{\mathcal{T}}_{3}^{-1}\cdot\big(\hat{a}_{\boldsymbol{k}}^{(s)}(\tau_{0})-\text{i}\text{e}^{-\text{i}\varphi_{k}^{(s)}(\tau)}\tan(r_{k}^{(s)}(\tau))\hat{a}_{-\boldsymbol{k}}^{(s)\dagger}(\tau_{0})\big)\cdot\hat{\mathcal{T}}_{3}\hat{\mathcal{T}}_{2}\hat{\mathcal{T}}_{1}\\
\nonumber
&=\hat{\mathcal{T}}_{1}^{-1}\hat{\mathcal{T}}_{2}^{-1}\cdot\big(\frac{1}{\cos(r_{k}^{(s)}(\tau))}\hat{a}_{\boldsymbol{k}}^{(s)}(\tau_{0})-\text{i}\text{e}^{-\text{i}\varphi_{k}^{(s)}(\tau)}\sin(r_{k}^{(s)}(\tau))\hat{a}_{-\boldsymbol{k}}^{(s)\dagger}(\tau_{0})\big)\cdot\hat{\mathcal{T}}_{2}\hat{\mathcal{T}}_{1}\\
\nonumber
&=\hat{\mathcal{T}}_{1}^{-1}\cdot\big(\cos(r_{k}^{(s)}(\tau))\hat{a}_{\boldsymbol{k}}^{(s)}(\tau_{0})-\text{i}\text{e}^{-\text{i}\varphi_{k}^{(s)}(\tau)}\sin(r_{k}^{(s)}(\tau))\hat{a}_{-\boldsymbol{k}}^{(s)\dagger}(\tau_{0})\big)\cdot\hat{\mathcal{T}}_{1}\\
\label{Task2BogoTransformOPE}
&=\cos(r_{k}^{(s)}(\tau))\text{e}^{\text{i}\omega_{k}^{(s)}(\tau)}\hat{a}_{\boldsymbol{k}}^{(s)}(\tau_{0})-\text{i}\sin(r_{k}^{(s)}(\tau))\text{e}^{-\text{i}(\omega_{k}^{(s)}(\tau)+\varphi_{k}^{(s)}(\tau))}\hat{a}_{-\boldsymbol{k}}^{(s)\dagger}(\tau_{0}),
\end{align}where the relation $\text{e}^{\hat{X}}\hat{Y}\text{e}^{-\hat{X}}=\hat{Y}+[\hat{X},\hat{Y}]+\frac{1}{2!}[\hat{X},[\hat{X},\hat{Y}]]+\frac{1}{3!}[\hat{X},[\hat{X},[\hat{X},\hat{Y}]]]+\dots$ need to be used frequently. The conjugate transformation follows analogously:
\begin{align}
\label{Task2BogoTransformOPEv1}
&\hat{a}_{-\boldsymbol{k}}^{(s)\dagger}(\tau)=\cos(r_{k}^{(s)}(\tau))\text{e}^{-\text{i}\omega_{k}^{(s)}(\tau)}\hat{a}_{-\boldsymbol{k}}^{(s)\dagger}(\tau_{0})+\text{i}\sin(r_{k}^{(s)}(\tau))\text{e}^{\text{i}(\omega_{k}^{(s)}(\tau)+\varphi_{k}^{(s)}(\tau))}\hat{a}_{\boldsymbol{k}}^{(s)}(\tau_{0}).
\end{align}Equations~\eqref{Task2BogoTransformOPE} and
\eqref{Task2BogoTransformOPEv1} preserve the canonical anticommutation
relations and constitute the fermionic Bogoliubov transformation.  We use the
following normalization convention for the action of creation and
annihilation operators on occupation-number states:
\begin{align}
\nonumber
&\hat{a}_{\pm\boldsymbol{k}}^{\dagger(s^{\prime})}(\tau_{0})\vert0_{\pm\boldsymbol{q}}^{(s)}\rangle=\delta^{s^{\prime}s}\delta_{\boldsymbol{k},\boldsymbol{q}}(2\pi)^{3/2}\vert1_{\pm\boldsymbol{q}}^{(s)}\rangle,\\
\nonumber
&\hat{a}_{\pm\boldsymbol{k}}^{(s^{\prime})}(\tau_{0})\vert1_{\pm\boldsymbol{q}}^{(s)}\rangle=\delta^{s^{\prime}s}\delta_{\boldsymbol{k},\boldsymbol{q}}(2\pi)^{3/2}\vert0_{\pm\boldsymbol{q}}^{(s)}\rangle,\\
\nonumber
&\hat{a}_{\pm\boldsymbol{k}}^{(s^{\prime})}(\tau_{0})\vert0_{\pm\boldsymbol{q}}^{(s)}\rangle=\hat{a}_{\pm\boldsymbol{k}}^{(s^{\prime})\dagger}(\tau_{0})\vert1_{\pm\boldsymbol{q}}^{(s)}\rangle=0.
\end{align}Finally, matching Eqs.~\eqref{Task2BogoTransformOPE} and
\eqref{Task2BogoTransformOPEv1} to the Hamiltonian-diagonalization result in
Eq.~\eqref{StandardBogoTransCreatAnnihi} determines the squeezing and phase
parameters in terms of $\mathcal A_k^{(s)}$, $\mathcal B_k^{(s)}$, and
$\omega_k^{(s)}$:
\begin{align}
\label{MatchSqueezingToHamiltonian}
&\cos(r_{k}^{(s)}(\tau))\text{e}^{\text{i}\omega_{k}^{(s)}(\tau)}=\frac{\mathcal{B}_{k}^{(s)}}{\sqrt{2\omega_{k}^{(s)}}\sqrt{\omega_{k}^{(s)}-\mathcal{A}_{k}^{(s)}}}~,~\sin(r_{k}^{(s)}(\tau))\text{e}^{-\text{i}(\omega_{k}^{(s)}(\tau)+\varphi_{k}^{(s)}(\tau))}=\frac{\text{i}\sqrt{\omega_{k}^{(s)}-\mathcal{A}_{k}^{(s)}}}{\sqrt{2\omega_{k}^{(s)}}}.
\end{align}Taking the modulus squared of the second relation in
Eq.~\eqref{MatchSqueezingToHamiltonian} gives
$\mathcal N_k^{(s)}=\sin^2r_k^{(s)}$, consistently with
Eq.~\eqref{ParticleNumberFromBogoliubov}.  The complex phases determine the
remaining squeezing and rotation parameters.  These relations lead directly
to Eqs.~\eqref{SqueezeAmplitude} and \eqref{SqueezeRotationAngles} in the main
text and connect the mode-function description, the instantaneous Bogoliubov
basis, and the fermionic two-mode state used in the quantum-information
analysis of Sec.~\ref{QuanInforMeasure}.
\end{widetext}

\bibliographystyle{unsrt}
\bibliography{bibliography}

\end{document}